\pdfoutput=1 
\documentclass[cernpreprint,english]{na61doc}
\usepackage[utf8]{inputenc}
\usepackage[T1]{fontenc}

\usepackage{lineno}
\usepackage{mathptmx} 
\usepackage{enumerate}
\usepackage{mathtools} 
\usepackage{amssymb}
\usepackage{placeins}
\usepackage{appendix}
\usepackage{color}
\usepackage{tabularx}
\usepackage{multirow}
\usepackage[colorinlistoftodos]{todonotes}

\usepackage{subcaption}
\usepackage{booktabs}
\usepackage{url}
\usepackage{cite}
\usepackage{caption}

\usepackage{hyperref}

\usepackage{svg}
\usepackage{bbding}
\usepackage{tikz}
\usetikzlibrary{patterns}
\usetikzlibrary{plotmarks}
\DeclareUnicodeCharacter{2212}{-}

\usepackage{diagbox}

\newcommand{\eV}{\ensuremath{\mbox{e\kern-0.1em V}}\xspace}
\newcommand{\GeV}{\ensuremath{\mbox{Ge\kern-0.1em V}}\xspace}
\newcommand{\MeV}{\ensuremath{\mbox{Me\kern-0.1em V}}\xspace}
\newcommand{\GeVc}{\ensuremath{\mbox{Ge\kern-0.1em V}\!/\!c}\xspace}
\newcommand{\GeVcc}{\ensuremath{\mbox{Ge\kern-0.1em V}\!/\!c^2}\xspace}
\newcommand{\AGeV}{\ensuremath{A\,\mbox{Ge\kern-0.1em V}}\xspace}
\newcommand{\AGeVc}{\ensuremath{A\,\mbox{Ge\kern-0.1em V}\!/\!c}\xspace}
\newcommand{\MeVc}{\ensuremath{\mbox{Me\kern-0.1em V}/c}\xspace}

\newcommand{\cm}{\ensuremath{\mbox{cm}}\xspace}

\newcommand{\dd}{\ensuremath{{\textrm d}}\xspace}
\newcommand{\dedx}{\ensuremath{\dd E/\dd x}\xspace}

\newcommand{\pt}{\ensuremath{p_{T}}\xspace}

\newcommand{\GeantThree}{{\scshape Geant3}\xspace}

\newcommand{\EposLong}{{\scshape Epos1.99}\xspace}

\newcommand{\NASixtyOne}{NA61\slash SHINE\xspace}
\newcommand{\CernVM}{\textsc{Cern\-\kern-0.05emVM}\xspace}

\newcommand{\Kshort}{$K^0_S$ }

\newcommand{\coordinate}[1]{{\fontfamily{lmss}\selectfont#1}}

\ShineTitle{\Kshort meson production in inelastic \textit{p+p} interactions at 31, 40 and 80~\GeVc beam momentum measured by \NASixtyOne at the CERN SPS}

\PreprintIdNumber{CERN-EP-2024-051}

\ShineJournal{}	

\ShineAbstract{
Measurements of \Kshort meson production via its $\pi^{+} \pi^{-}$ decay mode in inelastic \textit{p+p} interactions at incident projectile momenta of 31, 40 and 80~\GeVc ($\sqrt{s_{NN}}=7.7, 8.8$ and $12.3$~\GeV, respectively) are presented. The data were recorded by the \NASixtyOne spectrometer at the CERN Super Proton Synchrotron. Double-differential distributions were obtained in transverse momentum and rapidity. The mean multiplicities of \Kshort mesons were determined to be $(5.95 \pm 0.19 (stat) \pm 0.22 (sys)) \times 10^{-2}$ at 31~\GeVc, $(7.61 \pm 0.13 (stat) \pm 0.31 (sys)) \times 10^{-2}$ at 40~\GeVc and $(11.58 \pm 0.12 (stat) \pm 0.37 (sys)) \times 10^{-2}$ at 80~\GeVc. The results on $K^{0}_{S}$ production are compared with model calculations (\EposLong, SMASH 2.0 and PHSD) as well as with published data from other experiments. 
}

\begin{document}

\maketitle

\pagebreak

\small
\tableofcontents
\normalsize

\pagebreak

\newpage 

\bigskip

\noindent
\mbox{N.\;Abgrall\href{https://orcid.org/0009-0005-0777-8661}{\includegraphics[height=1.7ex]{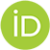}}\textsuperscript{\,29}},
\mbox{H.\;Adhikary\href{https://orcid.org/0000-0002-5746-1268}{\includegraphics[height=1.7ex]{figures/orcid-logo.png}}\textsuperscript{\,13}},
\mbox{P.\;Adrich\href{https://orcid.org/0000-0002-7019-5451}{\includegraphics[height=1.7ex]{figures/orcid-logo.png}}\textsuperscript{\,15}},
\mbox{K.K.\;Allison\href{https://orcid.org/0000-0002-3494-9383}{\includegraphics[height=1.7ex]{figures/orcid-logo.png}}\textsuperscript{\,26}},
\mbox{N.\;Amin\href{https://orcid.org/0009-0004-7572-3817}{\includegraphics[height=1.7ex]{figures/orcid-logo.png}}\textsuperscript{\,5}},
\mbox{E.V.\;Andronov\href{https://orcid.org/0000-0003-0437-9292}{\includegraphics[height=1.7ex]{figures/orcid-logo.png}}\textsuperscript{\,22}},
\mbox{T.\;Anti\'ci\'c\href{https://orcid.org/0000-0002-6606-0191}{\includegraphics[height=1.7ex]{figures/orcid-logo.png}}\textsuperscript{\,3}},
\mbox{I.-C.\;Arsene\href{https://orcid.org/0000-0003-2316-9565}{\includegraphics[height=1.7ex]{figures/orcid-logo.png}}\textsuperscript{\,12}},
\mbox{M.\;Bajda\href{https://orcid.org/0009-0005-8859-1099}{\includegraphics[height=1.7ex]{figures/orcid-logo.png}}\textsuperscript{\,16}},
\mbox{Y.\;Balkova\href{https://orcid.org/0000-0002-6957-573X}{\includegraphics[height=1.7ex]{figures/orcid-logo.png}}\textsuperscript{\,18}},
\mbox{M.\;Baszczyk\href{https://orcid.org/0000-0002-2595-0104}{\includegraphics[height=1.7ex]{figures/orcid-logo.png}}\textsuperscript{\,17}},
\mbox{D.\;Battaglia\href{https://orcid.org/0000-0002-5283-0992}{\includegraphics[height=1.7ex]{figures/orcid-logo.png}}\textsuperscript{\,25}},
\mbox{A.\;Bazgir\href{https://orcid.org/0000-0003-0358-0576}{\includegraphics[height=1.7ex]{figures/orcid-logo.png}}\textsuperscript{\,13}},
\mbox{S.\;Bhosale\href{https://orcid.org/0000-0001-5709-4747}{\includegraphics[height=1.7ex]{figures/orcid-logo.png}}\textsuperscript{\,14}},
\mbox{M.\;Bielewicz\href{https://orcid.org/0000-0001-8267-4874}{\includegraphics[height=1.7ex]{figures/orcid-logo.png}}\textsuperscript{\,15}},
\mbox{A.\;Blondel\href{https://orcid.org/0000-0002-1597-8859}{\includegraphics[height=1.7ex]{figures/orcid-logo.png}}\textsuperscript{\,4}},
\mbox{M.\;Bogomilov\href{https://orcid.org/0000-0001-7738-2041}{\includegraphics[height=1.7ex]{figures/orcid-logo.png}}\textsuperscript{\,2}},
\mbox{Y.\;Bondar\href{https://orcid.org/0000-0003-2773-9668}{\includegraphics[height=1.7ex]{figures/orcid-logo.png}}\textsuperscript{\,13}},
\mbox{N.\;Bostan\href{https://orcid.org/0000-0002-1129-4345}{\includegraphics[height=1.7ex]{figures/orcid-logo.png}}\textsuperscript{\,25}},
\mbox{A.\;Brandin\textsuperscript{\,22}},
\mbox{A.\;Bravar\href{https://orcid.org/0000-0002-1134-1527}{\includegraphics[height=1.7ex]{figures/orcid-logo.png}}\textsuperscript{\,29}},\\
\mbox{W.\;Bryli\'nski\href{https://orcid.org/0000-0002-3457-6601}{\includegraphics[height=1.7ex]{figures/orcid-logo.png}}\textsuperscript{\,21}},
\mbox{J.\;Brzychczyk\href{https://orcid.org/0000-0001-5320-6748}{\includegraphics[height=1.7ex]{figures/orcid-logo.png}}\textsuperscript{\,16}},
\mbox{M.\;Buryakov\href{https://orcid.org/0009-0008-2394-4967}{\includegraphics[height=1.7ex]{figures/orcid-logo.png}}\textsuperscript{\,22}},
\mbox{A.F.\;Camino\textsuperscript{\,28}},
\mbox{M.\;\'Cirkovi\'c\href{https://orcid.org/0000-0002-4420-9688}{\includegraphics[height=1.7ex]{figures/orcid-logo.png}}\textsuperscript{\,23}},
\mbox{M.\;Csan\'ad\href{https://orcid.org/0000-0002-3154-6925}{\includegraphics[height=1.7ex]{figures/orcid-logo.png}}\textsuperscript{\,8}},
\mbox{J.\;Cybowska\href{https://orcid.org/0000-0003-2568-3664}{\includegraphics[height=1.7ex]{figures/orcid-logo.png}}\textsuperscript{\,21}},
\mbox{T.\;Czopowicz\href{https://orcid.org/0000-0003-1908-2977}{\includegraphics[height=1.7ex]{figures/orcid-logo.png}}\textsuperscript{\,13}},
\mbox{C.\;Dalmazzone\href{https://orcid.org/0000-0001-6945-5845}{\includegraphics[height=1.7ex]{figures/orcid-logo.png}}\textsuperscript{\,4}},
\mbox{N.\;Davis\href{https://orcid.org/0000-0003-3047-6854}{\includegraphics[height=1.7ex]{figures/orcid-logo.png}}\textsuperscript{\,14}},
\mbox{A.\;Dmitriev\href{https://orcid.org/0000-0001-7853-0173}{\includegraphics[height=1.7ex]{figures/orcid-logo.png}}\textsuperscript{\,22}},
\mbox{P.~von\;Doetinchem\href{https://orcid.org/0000-0002-7801-3376}{\includegraphics[height=1.7ex]{figures/orcid-logo.png}}\textsuperscript{\,27}},
\mbox{W.\;Dominik\href{https://orcid.org/0000-0001-7444-9239}{\includegraphics[height=1.7ex]{figures/orcid-logo.png}}\textsuperscript{\,19}},
\mbox{P.\;Dorosz\href{https://orcid.org/0000-0002-8884-0981}{\includegraphics[height=1.7ex]{figures/orcid-logo.png}}\textsuperscript{\,17}},
\mbox{J.\;Dumarchez\href{https://orcid.org/0000-0002-9243-4425}{\includegraphics[height=1.7ex]{figures/orcid-logo.png}}\textsuperscript{\,4}},
\mbox{R.\;Engel\href{https://orcid.org/0000-0003-2924-8889}{\includegraphics[height=1.7ex]{figures/orcid-logo.png}}\textsuperscript{\,5}},
\mbox{G.A.\;Feofilov\href{https://orcid.org/0000-0003-3700-8623}{\includegraphics[height=1.7ex]{figures/orcid-logo.png}}\textsuperscript{\,22}},
\mbox{L.\;Fields\href{https://orcid.org/0000-0001-8281-3686}{\includegraphics[height=1.7ex]{figures/orcid-logo.png}}\textsuperscript{\,25}},
\mbox{Z.\;Fodor\href{https://orcid.org/0000-0003-2519-5687}{\includegraphics[height=1.7ex]{figures/orcid-logo.png}}\textsuperscript{\,7,20}},
\mbox{M.\;Friend\href{https://orcid.org/0000-0003-4660-4670}{\includegraphics[height=1.7ex]{figures/orcid-logo.png}}\textsuperscript{\,9}},
\mbox{M.\;Ga\'zdzicki\href{https://orcid.org/0000-0002-6114-8223}{\includegraphics[height=1.7ex]{figures/orcid-logo.png}}\textsuperscript{\,13,6}},
\mbox{O.\;Golosov\href{https://orcid.org/0000-0001-6562-2925}{\includegraphics[height=1.7ex]{figures/orcid-logo.png}}\textsuperscript{\,22}},
\mbox{V.\;Golovatyuk\href{https://orcid.org/0009-0006-5201-0990}{\includegraphics[height=1.7ex]{figures/orcid-logo.png}}\textsuperscript{\,22}},
\mbox{M.\;Golubeva\href{https://orcid.org/0009-0003-4756-2449}{\includegraphics[height=1.7ex]{figures/orcid-logo.png}}\textsuperscript{\,22}},
\mbox{K.\;Grebieszkow\href{https://orcid.org/0000-0002-6754-9554}{\includegraphics[height=1.7ex]{figures/orcid-logo.png}}\textsuperscript{\,21}},
\mbox{F.\;Guber\href{https://orcid.org/0000-0001-8790-3218}{\includegraphics[height=1.7ex]{figures/orcid-logo.png}}\textsuperscript{\,22}},
\mbox{A.\;Haesler\textsuperscript{\,29}},
\mbox{S.N.\;Igolkin\textsuperscript{\,22}},
\mbox{S.\;Ilieva\href{https://orcid.org/0000-0001-9204-2563}{\includegraphics[height=1.7ex]{figures/orcid-logo.png}}\textsuperscript{\,2}},
\mbox{A.\;Ivashkin\href{https://orcid.org/0000-0003-4595-5866}{\includegraphics[height=1.7ex]{figures/orcid-logo.png}}\textsuperscript{\,22}},
\mbox{A.\;Izvestnyy\href{https://orcid.org/0009-0009-1305-7309}{\includegraphics[height=1.7ex]{figures/orcid-logo.png}}\textsuperscript{\,22}},
\mbox{K.\;Kadija\textsuperscript{\,3}},
\mbox{N.\;Kargin\textsuperscript{\,22}},
\mbox{N.\;Karpushkin\href{https://orcid.org/0000-0001-5513-9331}{\includegraphics[height=1.7ex]{figures/orcid-logo.png}}\textsuperscript{\,22}},
\mbox{E.\;Kashirin\href{https://orcid.org/0000-0001-6062-7997}{\includegraphics[height=1.7ex]{figures/orcid-logo.png}}\textsuperscript{\,22}},
\mbox{M.\;Kie{\l}bowicz\href{https://orcid.org/0000-0002-4403-9201}{\includegraphics[height=1.7ex]{figures/orcid-logo.png}}\textsuperscript{\,14}},
\mbox{V.A.\;Kireyeu\href{https://orcid.org/0000-0002-5630-9264}{\includegraphics[height=1.7ex]{figures/orcid-logo.png}}\textsuperscript{\,22}},
\mbox{H.\;Kitagawa\textsuperscript{\,10}},
\mbox{R.\;Kolesnikov\href{https://orcid.org/0009-0006-4224-1058}{\includegraphics[height=1.7ex]{figures/orcid-logo.png}}\textsuperscript{\,22}},
\mbox{D.\;Kolev\href{https://orcid.org/0000-0002-9203-4739}{\includegraphics[height=1.7ex]{figures/orcid-logo.png}}\textsuperscript{\,2}},
\mbox{A.\;Korzenev\href{https://orcid.org/0000-0003-2107-4415}{\includegraphics[height=1.7ex]{figures/orcid-logo.png}}\textsuperscript{\,29}},\mbox{Y.\;Koshio\textsuperscript{\,10}},
\mbox{V.N.\;Kovalenko\href{https://orcid.org/0000-0001-6012-6615}{\includegraphics[height=1.7ex]{figures/orcid-logo.png}}\textsuperscript{\,22}},
\mbox{S.\;Kowalski\href{https://orcid.org/0000-0001-9888-4008}{\includegraphics[height=1.7ex]{figures/orcid-logo.png}}\textsuperscript{\,18}},
\mbox{B.\;Koz{\l}owski\href{https://orcid.org/0000-0001-8442-2320}{\includegraphics[height=1.7ex]{figures/orcid-logo.png}}\textsuperscript{\,21}},
\mbox{A.\;Krasnoperov\href{https://orcid.org/0000-0002-1425-2861}{\includegraphics[height=1.7ex]{figures/orcid-logo.png}}\textsuperscript{\,22}},
\mbox{W.\;Kucewicz\href{https://orcid.org/0000-0002-2073-711X}{\includegraphics[height=1.7ex]{figures/orcid-logo.png}}\textsuperscript{\,17}},
\mbox{M.\;Kuchowicz\href{https://orcid.org/0000-0003-3174-585X}{\includegraphics[height=1.7ex]{figures/orcid-logo.png}}\textsuperscript{\,20}},
\mbox{M.\;Kuich\href{https://orcid.org/0000-0002-6507-8699}{\includegraphics[height=1.7ex]{figures/orcid-logo.png}}\textsuperscript{\,19}},
\mbox{A.\;Kurepin\href{https://orcid.org/0000-0002-1851-4136}{\includegraphics[height=1.7ex]{figures/orcid-logo.png}}\textsuperscript{\,22}},
\mbox{A.\;L\'aszl\'o\href{https://orcid.org/0000-0003-2712-6968}{\includegraphics[height=1.7ex]{figures/orcid-logo.png}}\textsuperscript{\,7}},
\mbox{M.\;Lewicki\href{https://orcid.org/0000-0002-8972-3066}{\includegraphics[height=1.7ex]{figures/orcid-logo.png}}\textsuperscript{\,20}},
\mbox{G.\;Lykasov\href{https://orcid.org/0000-0002-1544-6959}{\includegraphics[height=1.7ex]{figures/orcid-logo.png}}\textsuperscript{\,22}},
\mbox{V.V.\;Lyubushkin\href{https://orcid.org/0000-0003-0136-233X}{\includegraphics[height=1.7ex]{figures/orcid-logo.png}}\textsuperscript{\,22}},
\mbox{M.\;Ma\'ckowiak-Paw{\l}owska\href{https://orcid.org/0000-0003-3954-6329}{\includegraphics[height=1.7ex]{figures/orcid-logo.png}}\textsuperscript{\,21}},
\mbox{Z.\;Majka\href{https://orcid.org/0000-0003-3064-6577}{\includegraphics[height=1.7ex]{figures/orcid-logo.png}}\textsuperscript{\,16}},
\mbox{A.\;Makhnev\href{https://orcid.org/0009-0002-9745-1897}{\includegraphics[height=1.7ex]{figures/orcid-logo.png}}\textsuperscript{\,22}},
\mbox{B.\;Maksiak\href{https://orcid.org/0000-0002-7950-2307}{\includegraphics[height=1.7ex]{figures/orcid-logo.png}}\textsuperscript{\,15}},
\mbox{A.I.\;Malakhov\href{https://orcid.org/0000-0001-8569-8409}{\includegraphics[height=1.7ex]{figures/orcid-logo.png}}\textsuperscript{\,22}},
\mbox{A.\;Marcinek\href{https://orcid.org/0000-0001-9922-743X}{\includegraphics[height=1.7ex]{figures/orcid-logo.png}}\textsuperscript{\,14}},
\mbox{A.D.\;Marino\href{https://orcid.org/0000-0002-1709-538X}{\includegraphics[height=1.7ex]{figures/orcid-logo.png}}\textsuperscript{\,26}},
\mbox{H.-J.\;Mathes\href{https://orcid.org/0000-0002-0680-040X}{\includegraphics[height=1.7ex]{figures/orcid-logo.png}}\textsuperscript{\,5}},
\mbox{T.\;Matulewicz\href{https://orcid.org/0000-0003-2098-1216}{\includegraphics[height=1.7ex]{figures/orcid-logo.png}}\textsuperscript{\,19}},
\mbox{V.\;Matveev\href{https://orcid.org/0000-0002-2745-5908}{\includegraphics[height=1.7ex]{figures/orcid-logo.png}}\textsuperscript{\,22}},
\mbox{G.L.\;Melkumov\href{https://orcid.org/0009-0004-2074-6755}{\includegraphics[height=1.7ex]{figures/orcid-logo.png}}\textsuperscript{\,22}},
\mbox{A.\;Merzlaya\href{https://orcid.org/0000-0002-6553-2783}{\includegraphics[height=1.7ex]{figures/orcid-logo.png}}\textsuperscript{\,12}},
\mbox{{\L}.\;Mik\href{https://orcid.org/0000-0003-2712-6861}{\includegraphics[height=1.7ex]{figures/orcid-logo.png}}\textsuperscript{\,17}},
\mbox{A.\;Morawiec\href{https://orcid.org/0009-0001-9845-4005}{\includegraphics[height=1.7ex]{figures/orcid-logo.png}}\textsuperscript{\,16}},
\mbox{S.\;Morozov\href{https://orcid.org/0000-0002-6748-7277}{\includegraphics[height=1.7ex]{figures/orcid-logo.png}}\textsuperscript{\,22}},
\mbox{Y.\;Nagai\href{https://orcid.org/0000-0002-1792-5005}{\includegraphics[height=1.7ex]{figures/orcid-logo.png}}\textsuperscript{\,8}},
\mbox{T.\;Nakadaira\href{https://orcid.org/0000-0003-4327-7598}{\includegraphics[height=1.7ex]{figures/orcid-logo.png}}\textsuperscript{\,9}},
\mbox{M.\;Naskr\k{e}t\href{https://orcid.org/0000-0002-5634-6639}{\includegraphics[height=1.7ex]{figures/orcid-logo.png}}\textsuperscript{\,20}},
\mbox{S.\;Nishimori\href{https://orcid.org/~0000-0002-1820-0938}{\includegraphics[height=1.7ex]{figures/orcid-logo.png}}\textsuperscript{\,9}},
\mbox{V.\;Ozvenchuk\href{https://orcid.org/0000-0002-7821-7109}{\includegraphics[height=1.7ex]{figures/orcid-logo.png}}\textsuperscript{\,14}},
\mbox{O.\;Panova\href{https://orcid.org/0000-0001-5039-7788}{\includegraphics[height=1.7ex]{figures/orcid-logo.png}}\textsuperscript{\,13}},
\mbox{V.\;Paolone\href{https://orcid.org/0000-0003-2162-0957}{\includegraphics[height=1.7ex]{figures/orcid-logo.png}}\textsuperscript{\,28}},
\mbox{O.\;Petukhov\href{https://orcid.org/0000-0002-8872-8324}{\includegraphics[height=1.7ex]{figures/orcid-logo.png}}\textsuperscript{\,22}},
\mbox{I.\;Pidhurskyi\href{https://orcid.org/0000-0001-9916-9436}{\includegraphics[height=1.7ex]{figures/orcid-logo.png}}\textsuperscript{\,13,6}},
\mbox{R.\;P{\l}aneta\href{https://orcid.org/0000-0001-8007-8577}{\includegraphics[height=1.7ex]{figures/orcid-logo.png}}\textsuperscript{\,16}},
\mbox{P.\;Podlaski\href{https://orcid.org/0000-0002-0232-9841}{\includegraphics[height=1.7ex]{figures/orcid-logo.png}}\textsuperscript{\,19}},
\mbox{B.A.\;Popov\href{https://orcid.org/0000-0001-5416-9301}{\includegraphics[height=1.7ex]{figures/orcid-logo.png}}\textsuperscript{\,22,4}},
\mbox{B.\;P\'orfy\href{https://orcid.org/0000-0001-5724-9737}{\includegraphics[height=1.7ex]{figures/orcid-logo.png}}\textsuperscript{\,7,8}},
\mbox{M.\;Posiada{\l}a-Zezula\href{https://orcid.org/0000-0002-5154-5348}{\includegraphics[height=1.7ex]{figures/orcid-logo.png}}\textsuperscript{\,19}},
\mbox{D.S.\;Prokhorova\href{https://orcid.org/0000-0003-3726-9196}{\includegraphics[height=1.7ex]{figures/orcid-logo.png}}\textsuperscript{\,22}},
\mbox{D.\;Pszczel\href{https://orcid.org/0000-0002-4697-6688}{\includegraphics[height=1.7ex]{figures/orcid-logo.png}}\textsuperscript{\,15}},
\mbox{S.\;Pu{\l}awski\href{https://orcid.org/0000-0003-1982-2787}{\includegraphics[height=1.7ex]{figures/orcid-logo.png}}\textsuperscript{\,18}},
\mbox{J.\;Puzovi\'c\textsuperscript{\,23}\textsuperscript{\dag}},
\mbox{R.\;Renfordt\href{https://orcid.org/0000-0002-5633-104X}{\includegraphics[height=1.7ex]{figures/orcid-logo.png}}\textsuperscript{\,18}},
\mbox{L.\;Ren\href{https://orcid.org/0000-0003-1709-7673}{\includegraphics[height=1.7ex]{figures/orcid-logo.png}}\textsuperscript{\,26}},
\mbox{V.Z.\;Reyna~Ortiz\href{https://orcid.org/0000-0002-7026-8198}{\includegraphics[height=1.7ex]{figures/orcid-logo.png}}\textsuperscript{\,13}},
\mbox{D.\;R\"ohrich\textsuperscript{\,11}},
\mbox{E.\;Rondio\href{https://orcid.org/0000-0002-2607-4820}{\includegraphics[height=1.7ex]{figures/orcid-logo.png}}\textsuperscript{\,15}},
\mbox{M.\;Roth\href{https://orcid.org/0000-0003-1281-4477}{\includegraphics[height=1.7ex]{figures/orcid-logo.png}}\textsuperscript{\,5}},
\mbox{{\L}.\;Rozp{\l}ochowski\href{https://orcid.org/0000-0003-3680-6738}{\includegraphics[height=1.7ex]{figures/orcid-logo.png}}\textsuperscript{\,14}},
\mbox{B.T.\;Rumberger\href{https://orcid.org/0000-0002-4867-945X}{\includegraphics[height=1.7ex]{figures/orcid-logo.png}}\textsuperscript{\,26}},
\mbox{M.\;Rumyantsev\href{https://orcid.org/0000-0001-8233-2030}{\includegraphics[height=1.7ex]{figures/orcid-logo.png}}\textsuperscript{\,22}},
\mbox{A.\;Rustamov\href{https://orcid.org/0000-0001-8678-6400}{\includegraphics[height=1.7ex]{figures/orcid-logo.png}}\textsuperscript{\,1,6}},
\mbox{M.\;Rybczynski\href{https://orcid.org/0000-0002-3638-3766}{\includegraphics[height=1.7ex]{figures/orcid-logo.png}}\textsuperscript{\,13}},
\mbox{A.\;Rybicki\href{https://orcid.org/0000-0003-3076-0505}{\includegraphics[height=1.7ex]{figures/orcid-logo.png}}\textsuperscript{\,14}},
\mbox{K.\;Sakashita\href{https://orcid.org/0000-0003-2602-7837}{\includegraphics[height=1.7ex]{figures/orcid-logo.png}}\textsuperscript{\,9}},
\mbox{K.\;Schmidt\href{https://orcid.org/0000-0002-0903-5790}{\includegraphics[height=1.7ex]{figures/orcid-logo.png}}\textsuperscript{\,18}},
\mbox{A.Yu.\;Seryakov\href{https://orcid.org/0000-0002-5759-5485}{\includegraphics[height=1.7ex]{figures/orcid-logo.png}}\textsuperscript{\,22}},
\mbox{P.\;Seyboth\href{https://orcid.org/0000-0002-4821-6105}{\includegraphics[height=1.7ex]{figures/orcid-logo.png}}\textsuperscript{\,13}},
\mbox{U.A.\;Shah\href{https://orcid.org/0000-0002-9315-1304}{\includegraphics[height=1.7ex]{figures/orcid-logo.png}}\textsuperscript{\,13}},
\mbox{Y.\;Shiraishi\textsuperscript{\,10}},
\mbox{A.\;Shukla\href{https://orcid.org/0000-0003-3839-7229}{\includegraphics[height=1.7ex]{figures/orcid-logo.png}}\textsuperscript{\,27}},
\mbox{M.\;S{\l}odkowski\href{https://orcid.org/0000-0003-0463-2753}{\includegraphics[height=1.7ex]{figures/orcid-logo.png}}\textsuperscript{\,21}},
\mbox{P.\;Staszel\href{https://orcid.org/0000-0003-4002-1626}{\includegraphics[height=1.7ex]{figures/orcid-logo.png}}\textsuperscript{\,16}},
\mbox{G.\;Stefanek\href{https://orcid.org/0000-0001-6656-9177}{\includegraphics[height=1.7ex]{figures/orcid-logo.png}}\textsuperscript{\,13}},
\mbox{J.\;Stepaniak\href{https://orcid.org/0000-0003-2064-9870}{\includegraphics[height=1.7ex]{figures/orcid-logo.png}}\textsuperscript{\,15}},
\mbox{M.\;Strikhanov\textsuperscript{\,22}},
\mbox{H.\;Str\"obele\textsuperscript{\,6}},
\mbox{T.\;\v{S}u\v{s}a\href{https://orcid.org/0000-0001-7430-2552}{\includegraphics[height=1.7ex]{figures/orcid-logo.png}}\textsuperscript{\,3}},
\mbox{L.\;Swiderski\href{https://orcid.org/0000-0001-5857-2085}{\includegraphics[height=1.7ex]{figures/orcid-logo.png}}\textsuperscript{\,15}},
\mbox{J.\;Szewi\'nski\href{https://orcid.org/0000-0003-2981-9303}{\includegraphics[height=1.7ex]{figures/orcid-logo.png}}\textsuperscript{\,15}},
\mbox{R.\;Szukiewicz\href{https://orcid.org/0000-0002-1291-4040}{\includegraphics[height=1.7ex]{figures/orcid-logo.png}}\textsuperscript{\,20}},
\mbox{A.\;Taranenko\href{https://orcid.org/0000-0003-1737-4474}{\includegraphics[height=1.7ex]{figures/orcid-logo.png}}\textsuperscript{\,22}},
\mbox{A.\;Tefelska\href{https://orcid.org/0000-0002-6069-4273}{\includegraphics[height=1.7ex]{figures/orcid-logo.png}}\textsuperscript{\,21}},
\mbox{D.\;Tefelski\href{https://orcid.org/0000-0003-0802-2290}{\includegraphics[height=1.7ex]{figures/orcid-logo.png}}\textsuperscript{\,21}},
\mbox{V.\;Tereshchenko\textsuperscript{\,22}},
\mbox{A.\;Toia\href{https://orcid.org/0000-0001-9567-3360}{\includegraphics[height=1.7ex]{figures/orcid-logo.png}}\textsuperscript{\,6}},
\mbox{R.\;Tsenov\href{https://orcid.org/0000-0002-1330-8640}{\includegraphics[height=1.7ex]{figures/orcid-logo.png}}\textsuperscript{\,2}},
\mbox{L.\;Turko\href{https://orcid.org/0000-0002-5474-8650}{\includegraphics[height=1.7ex]{figures/orcid-logo.png}}\textsuperscript{\,20}},
\mbox{T.S.\;Tveter\href{https://orcid.org/0009-0003-7140-8644}{\includegraphics[height=1.7ex]{figures/orcid-logo.png}}\textsuperscript{\,12}},
\mbox{M.\;Unger\href{https://orcid.org/0000-0002-7651-0272~}{\includegraphics[height=1.7ex]{figures/orcid-logo.png}}\textsuperscript{\,5}},
\mbox{M.\;Urbaniak\href{https://orcid.org/0000-0002-9768-030X}{\includegraphics[height=1.7ex]{figures/orcid-logo.png}}\textsuperscript{\,18}},
\mbox{F.F.\;Valiev\href{https://orcid.org/0000-0001-5130-5603}{\includegraphics[height=1.7ex]{figures/orcid-logo.png}}\textsuperscript{\,22}},
\mbox{D.\;Veberi\v{c}\href{https://orcid.org/0000-0003-2683-1526}{\includegraphics[height=1.7ex]{figures/orcid-logo.png}}\textsuperscript{\,5}},
\mbox{V.V.\;Vechernin\href{https://orcid.org/0000-0003-1458-8055}{\includegraphics[height=1.7ex]{figures/orcid-logo.png}}\textsuperscript{\,22}},
\mbox{V.\;Volkov\href{https://orcid.org/0000-0002-4785-7517}{\includegraphics[height=1.7ex]{figures/orcid-logo.png}}\textsuperscript{\,22}},
\mbox{A.\;Wickremasinghe\href{https://orcid.org/0000-0002-5325-0455}{\includegraphics[height=1.7ex]{figures/orcid-logo.png}}\textsuperscript{\,24}},
\mbox{K.\;W\'ojcik\href{https://orcid.org/0000-0002-8315-9281}{\includegraphics[height=1.7ex]{figures/orcid-logo.png}}\textsuperscript{\,18}},
\mbox{O.\;Wyszy\'nski\href{https://orcid.org/0000-0002-6652-0450}{\includegraphics[height=1.7ex]{figures/orcid-logo.png}}\textsuperscript{\,13}},
\mbox{A.\;Zaitsev\href{https://orcid.org/0000-0003-4711-9925}{\includegraphics[height=1.7ex]{figures/orcid-logo.png}}\textsuperscript{\,22}},
\mbox{E.D.\;Zimmerman\href{https://orcid.org/0000-0002-6394-6659}{\includegraphics[height=1.7ex]{figures/orcid-logo.png}}\textsuperscript{\,26}},
\mbox{A.\;Zviagina\href{https://orcid.org/0009-0007-5211-6493}{\includegraphics[height=1.7ex]{figures/orcid-logo.png}}\textsuperscript{\,22}}, and
\mbox{R.\;Zwaska\href{https://orcid.org/0000-0002-4889-5988}{\includegraphics[height=1.7ex]{figures/orcid-logo.png}}\textsuperscript{\,24}}
\\\rule{2cm}{.5pt}\\[-.5ex]\textit{\textsuperscript{\dag} \footnotesize deceased}

{\Large The \NASixtyOne Collaboration}


\noindent
\textsuperscript{1}~National Nuclear Research Center, Baku, Azerbaijan\\
\textsuperscript{2}~Faculty of Physics, University of Sofia, Sofia, Bulgaria\\
\textsuperscript{3}~Ru{\dj}er Bo\v{s}kovi\'c Institute, Zagreb, Croatia\\
\textsuperscript{4}~LPNHE, University of Paris VI and VII, Paris, France\\
\textsuperscript{5}~Karlsruhe Institute of Technology, Karlsruhe, Germany\\
\textsuperscript{6}~University of Frankfurt, Frankfurt, Germany\\
\textsuperscript{7}~Wigner Research Centre for Physics, Budapest, Hungary\\
\textsuperscript{8}~E\"otv\"os Lor\'and University, Budapest, Hungary\\
\textsuperscript{9}~Institute for Particle and Nuclear Studies, Tsukuba, Japan\\
\textsuperscript{10}~Okayama University, Japan\\
\textsuperscript{11}~University of Bergen, Bergen, Norway\\
\textsuperscript{12}~University of Oslo, Oslo, Norway\\
\textsuperscript{13}~Jan Kochanowski University, Kielce, Poland\\
\textsuperscript{14}~Institute of Nuclear Physics, Polish Academy of Sciences, Cracow, Poland\\
\textsuperscript{15}~National Centre for Nuclear Research, Warsaw, Poland\\
\textsuperscript{16}~Jagiellonian University, Cracow, Poland\\
\textsuperscript{17}~AGH - University of Science and Technology, Cracow, Poland\\
\textsuperscript{18}~University of Silesia, Katowice, Poland\\
\textsuperscript{19}~University of Warsaw, Warsaw, Poland\\
\textsuperscript{20}~University of Wroc{\l}aw,  Wroc{\l}aw, Poland\\
\textsuperscript{21}~Warsaw University of Technology, Warsaw, Poland\\
\textsuperscript{22}~Affiliated with an institution covered by a cooperation agreement with CERN\\
\textsuperscript{23}~University of Belgrade, Belgrade, Serbia\\
\textsuperscript{24}~Fermilab, Batavia, USA\\
\textsuperscript{25}~University of Notre Dame, Notre Dame, USA\\
\textsuperscript{26}~University of Colorado, Boulder, USA\\
\textsuperscript{27}~University of Hawaii at Manoa, Honolulu, USA\\
\textsuperscript{28}~University of Pittsburgh, Pittsburgh, USA\\
\textsuperscript{29}~University of Geneva, Geneva, Switzerland\footnote{No longer affiliated with the NA61/SHINE collaboration}\\


\section{Introduction}
\label{sec:intro}
The measurement of hadron production in proton-proton interactions plays a key role in understanding nucleus-nucleus collisions. In particular, it can shed some light on the creation process of Quark Gluon Plasma (QGP), on its properties, and on the characterization of the phase transition between hadronic matter and the QGP. One of the key signals of QGP creation is the enhanced production of $s$ and $\bar{s}$ quarks, carried mostly by kaons~\cite{Letessier:2002ony}. The experimental results indicate that the creation of the QGP starts in nucleus-nucleus collisions at centre of mass energies from 10 to 20 \GeV 
\cite{Gazdzicki:1998vd}, which is the realm of the \NASixtyOne experiment at CERN. To explore this region systematically \NASixtyOne studies observables indicative of the QGP by a two-dimensional scan in collision energy and nuclear mass number of the colliding nuclei. Since 2009, \NASixtyOne has collected data on \textit{p+p}, \textit{p}+Pb, Be+Be, Ar+Sc, Xe+La and Pb+Pb interactions in the beam momentum range from 13\textit{A} to 158\textit{A} \GeVc \cite{Gazdzicki:995681}. Results on neutral kaon spectra in \textit{p+p} at 158~\GeVc can be found in Ref.~\cite{NA61SHINE:2021iay}, while results on charged kaon spectra in \textit{p+p} at 31, 40, 80 and 158~\GeVc can be found in Ref.~\cite{NA61SHINE:2017fne}. In this paper, we present the results of \Kshort production in \textit{p+p} collisions at 31, 40, and 80 \GeVc, which will be compared with \Kshort production at 158 \GeVc \cite{NA61SHINE:2021iay} and constitute the baseline for the interpretation of results obtained in heavier systems collected with the \NASixtyOne detector. Thanks to high statistics, large acceptance, and good momentum resolution the results presented here have, in general, higher precision than previously published measurements at SPS energies.


The paper is organised as follows. In Sec.~\ref{sec:setup}, details of the \NASixtyOne detector system are presented. Section \ref{sec:method} is devoted to describing the analysis method. The results are shown in Sec.~\ref{sec:results}. In Sec.~\ref{sec:comparison}, they are compared to published world data and model calculations. Section~\ref{sec:summary} closes the paper with a summary and outlook.

The following units, variables and definitions are used in this paper. The particle mass and energy are presented in \GeV, while particle momentum is shown in \GeVc. The particle rapidity $y$ is calculated in the proton-proton collision center of mass system (cms), $y=0.5 \cdot ln[(E+cp_L)/(E-cp_L)]$, where $E$ and $p_L$ are the particle energy and longitudinal momentum. The transverse component of the momentum is denoted as $p_T$. The momentum in the laboratory frame is denoted $p_{lab}$ and the collision energy per nucleon pair in the centre of mass by $\sqrt{s_{NN}}$.

\FloatBarrier
\section{Experimental setup}
\label{sec:setup}
\begin{figure*}[ht]
  \centering
  \includegraphics[width=0.8\textwidth]{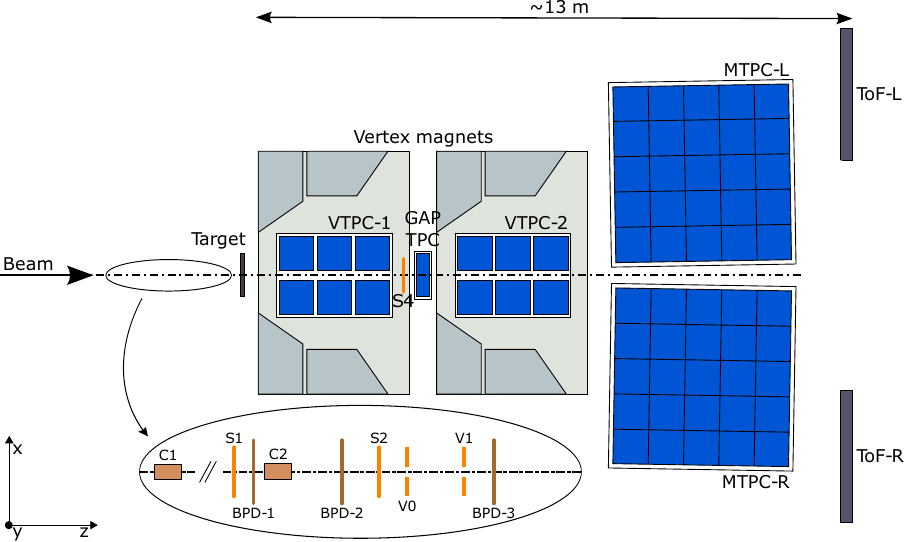}
  \caption[]{
    (Color online) The schematic layout of the NA61/SHINE experiment at the CERN SPS during \textit{p+p} data taking (horizontal cut, not to scale). The beam and trigger detector configuration used for data taking in 2009 is shown in the inset (see Refs.~\cite{Abgrall:2014xwa, NA61SHINE:2013tiv, Aduszkiewicz:2015jna} for a detailed description). The chosen coordinate system is drawn on the lower left: its origin lies in the middle of the VTPC-2 on the beam axis. }
  \label{fig:detector-setup}
\end{figure*}

The \NASixtyOne collaboration uses a large acceptance spectrometer located in the CERN North Area. The schematic layout of the \NASixtyOne detector during the \textit{p+p} data-taking is shown in Fig.~\ref{fig:detector-setup}. A detailed description of the experimental setup can be found in Ref.~\cite{Abgrall:2014xwa}, while the details on the simulation in describing the detector performance across different kinematic variables as well as its inefficiencies can be found in Ref. \cite{NA61SHINE:2013tiv}.

The main components of the \NASixtyOne spectrometer are four large-volume Time Projection Chambers (TPCs). Two of them, the vertex TPCs (\mbox{VTPC-1} and \mbox{VTPC-2}), are located in the magnetic fields of two super-conducting dipole magnets with a maximum combined bending power of 9~Tm, which corresponds to about 1.5~T and 1.1~T in the upstream and downstream magnets, respectively. This field strength was used for data taking at 158~\GeVc and scaled down in proportion to the lower beam momenta to obtain similar $y-p_T$ acceptance at all beam momenta. Two large main TPCs (\mbox{MTPC-L} and \mbox{MTPC-R}) and two walls of pixel Time-of-Flight (ToF-L/R) detectors are positioned symmetrically to the beamline downstream of the magnets. A GAP-TPC (GTPC) is placed between \mbox{VTPC-1} and \mbox{VTPC-2} directly on the beamline. It closes the gap between the beam axis and the sensitive volumes of the other TPCs. The TPCs are filled with Ar and CO$_2$ gas mixtures in proportions 90:10 for the VTPCs and 95:5 for the MTPCs. Particle identification in the TPCs is based on measurements of the specific energy loss (\dedx) in the chamber gas. Typical values for the momentum resolution are $\sigma(p)/p^2=7\times10^{-4}$~(\GeVc)$^{-1}$ for low-momentum tracks measured only in \mbox{VTPC-1} ($p\leq 8$~\GeVc) and $3\times10^{-3}$~(\GeVc)$^{-1}$ for tracks traversing the full detector up to and including the \mbox{MTPCs} ($p\geq 8$~\GeVc).

Secondary beams of positively charged hadrons at momenta of 31, 40 and 80~\GeVc were used to collect the data for the analysis presented in this paper. These beams were produced from a 400~\GeVc proton beam extracted from the SPS in a slow extraction mode with a flat-top of 10 seconds. The beam momentum and intensity were adjusted by appropriate settings of the H2 beam line magnet currents and collimators. Protons from the secondary hadron beam are identified by two Cherenkov counters, C1~\cite{Bovet:1982xf} and C2 (THC). The C1 counter, using a coincidence of six out of the eight photomultipliers placed radially along the Cherenkov ring, provides identification of protons, while the THC, operated at a pressure lower than the proton threshold, is used in anti-coincidence in the trigger logic. A selection based on the signals from the Cherenkov counters allowed to identify beam protons with a purity of about 99\%, as demonstrated by a measurement of the specific ionization energy loss \dedx of the beam particles by bending the 31~\GeVc beam into the TPCs using the full magnetic field strength ~\cite{Strabel:2011zz}.
A set of scintillation (S1, S2 and V0, V1) and beam position detectors (BPDs) upstream of the spectrometer provide timing reference, and position measurements of incoming beam particles. The trigger scintillation counter S4 placed downstream of the target has a diameter of 2 cm. It is used to trigger the readout whenever an incoming beam particle, which is registered upstream of the target, does not hit S4, which indicates that an interaction occurred in the target area.

A cylindrical target vessel of 20.29 cm length and 3 cm diameter was placed upstream of the entrance window of VTPC-1 (center of the target is at \coordinate{z} = -581~\cm in the NA61/SHINE coordinate system). The vessel was filled with liquid hydrogen corresponding to an interaction length of 2.8\%. The liquid hydrogen had a density of approximately 0.07 g/cm$^3$. Data were taken with the vessel filled with liquid hydrogen and being empty. Here, only events recorded with the target vessel filled with hydrogen were analyzed.



\FloatBarrier
\section{Analysis}\label{sec:method}

\subsection{Data sets}
\label{sec:data_sets}
The presented results on \Kshort production in inelastic \textit{p+p} interactions at $p_{beam}=31, 40$ and $80$~\GeVc are based on data recorded in 2009. Table~\ref{tab:datasets} summarizes basic information about data sets used in the analysis, the number of events selected by interaction trigger and the number of events after analysis cuts. The event numbers recorded with the interaction trigger were 2.85M, 4.37M and 3.80M, respectively. The drop in event numbers after cuts is caused mainly by BPD reconstruction inefficiencies and off-target interactions accepted by the trigger logic.  

\begin{table} [h]
\small
\centering
\begin{tabular}{|c|c|c|c|c|}
    \hline
    $p_{beam}$~(\GeVc) & $\sqrt{s_{NN}}$ (GeV) & Number of recorded events & Number of events after \\
     & & with interaction trigger & selection criteria \\
    \hline
    \hline 
    31 & 7.7 & 2.85 $\times 10^6$ & 0.83 $\times 10^6$ \\
    40 & 8.8 & 4.37 $\times 10^6$ & 1.24 $\times 10^6$ \\
    80 & 12.3 & 3.80 $\times 10^6$ & 1.48 $\times 10^6$ \\
    \hline
\end{tabular}
\caption{Data sets used for the analysis of \Kshort production. The beam momentum is denoted by $p_{beam}$, whereas $\sqrt{s_{NN}}$ is the energy available in the center-of-mass system for the nucleon pair. The events selection criteria are described in Sec.~\ref{s:event_selection}.}
\label{tab:datasets}
\end{table}

\subsection{Analysis method}
\label{sec:analysis_method}
The event vertex and the produced particle tracks were reconstructed using the standard \NASixtyOne software. Details of the track and vertex reconstruction procedures can be found in Refs.~\cite{NA61SHINE:2013tiv, Aduszkiewicz:2015jna, Aduszkiewicz:2016mww}. Detector parameters were optimized by a data-based calibration procedure, which also considered their time dependence; for details, see Refs.~\cite{NA61SHINE:2017fne, Abgrall:1113279}. The following section enumerates the criteria for selecting events, tracks and the \Kshort decay topology. Then, the simulation-based correction procedure is described and used to quantify the losses due to reconstruction inefficiencies and limited geometrical acceptance.

\subsection{Event selection}
\label{s:event_selection}
The criteria for selection of inelastic \textit{p+p} interactions are the following:

\begin{itemize}
    \item [(i)] An event was accepted by the trigger logic (see Refs.~\cite{NA61SHINE:2013tiv, Aduszkiewicz:2015jna}) as an interaction candidate event.
    \item [(ii)] No off-time beam particle was detected within a time window of $\pm$2 $\mu$s around the trigger particle.
    \item [(iii)] Beam particle trajectory was measured in at least three planes out of four of BPD-1 and BPD-2 and in both planes of BPD-3.
    \item [(iv)] The primary interaction vertex fit converged.
    \item [(v)] The \coordinate{z} position of the interaction vertex (fitted using the beam trajectory and TPC tracks) not farther away than 9~\cm from the center of the target vessel.
    \item [(vi)] Events with a single, well-measured, positively charged track with absolute momentum close to the beam momentum ($p>p_{beam}-1$~\GeVc) were rejected.
\end{itemize}

The background due to elastic interactions was removed via cuts (iv) and (vi). The contribution from off-target interactions was reduced by cut (v). The simulations corrected the losses of inelastic \textit{p+p} interactions due to the event selection procedure.

The numbers of events left after the selection criteria described in the text above are given in Table~\ref{tab:datasets}.

\subsection{Track and topology selection}
\label{s:track_selection}
Neutral strange particles are detected and measured using their weak decay into charged particles. The \Kshort decays into $\pi^+ + \pi^-$ with a branching ratio of 69.2\% ~\cite{PDG} are used here. The decay particles form the so-called $V^0$ topology. \Kshort decay candidates ($V^0$s) are obtained by pairing all positively with all negatively charged pion candidates. The tracks of the decay pions and the $V^0$ topology are subject to the following additional selection criteria:

\begin{itemize}
    \item [(i)] For each candidate track, the minimum number of measured clusters in VTPC-1 and VTPC-2 must be 15.
    \item [(ii)] All pion tracks must have a measured specific energy loss (\dedx) in the TPCs within $\pm 3\sigma$ around the nominal Bethe-Bloch value for charged pions. Here, $\sigma$ represents the typical standard deviation of a Gaussian fitted to the \dedx distribution of pions. Since only small variations of $\sigma$ were observed for different bins and beam momenta, a constant value $\sigma = 0.052$ is used \cite{NA61SHINE:2020czr}. This selection criterion applies only to experimental data, not MC-simulated events (see below).
    \item [(iii)] The distance |$\Delta$\coordinate{z}| between the \coordinate{z}-coordinates of the primary production and the \Kshort decay vertices is required to lie in the rapidity dependent range: |$\Delta$\coordinate{z}| > $e^{a+b \cdot y_{lab}}$, with $y_{lab}$ the rapidity in the laboratory and $a$ and $b$ constants which amount to 1.91 and 0.99 for the $p_{beam}=31$~\GeVc, 1.71 and 0.95 for $p_{beam}=40$~\GeVc, and 1.85 and 0.90 for $p_{beam}=80$~\GeVc data sets, respectively.
    \item [(iv)] The distance of closest approach (DCA) in the x and y directions of the straight line given by the \Kshort momentum vector in the laboratory and the primary vertex must be smaller than 0.25 \cm, with DCA given by
    $\sqrt{\left({b_x}/{2}\right)^2 + {b_y}^2}$.
    \item [(v)] The cosine of the angle between the $V^0$ and $\pi^+$ momentum vectors in the \Kshort rest frame has to be in the range: $-0.97 < cos\Theta^{*} < 0.85$.  
\end{itemize}

\begin{figure*}[h]
\centering
    \includegraphics[width=0.32\textwidth]{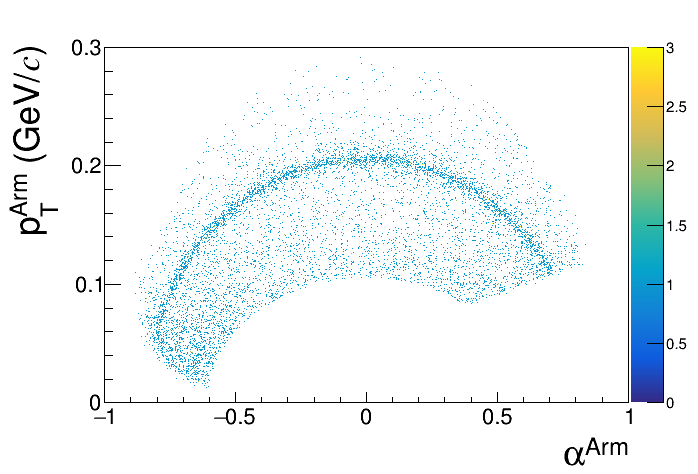}
    \includegraphics[width=0.32\textwidth]{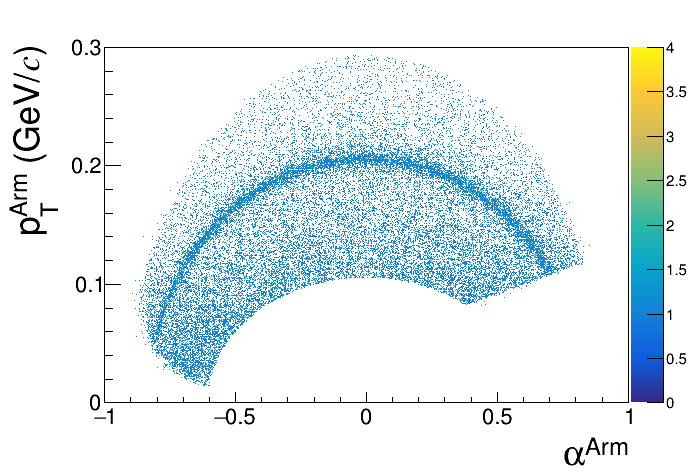} 
    \includegraphics[width=0.32\textwidth]{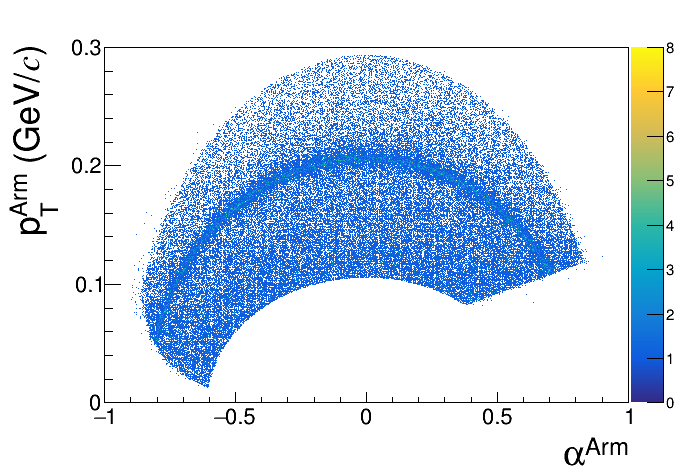} 
\caption[]{Armenteros-Podolanski plots of $V^0$ candidates after all track and topology selection criteria for $p_{beam}=31, 40$ and $80$~\GeVc from left to right. The boundaries on the plots' left and right sides result from using the $cos\Theta^{*}$ cut, while the upper and lower boundaries are shaped by selecting a certain invariant mass range.} 
\label{fig:APplots} 
\end{figure*}

The quality of the aforementioned track and topology selection criteria is illustrated in Fig.~\ref{fig:APplots}. The population of \Kshort decay candidates is shown as a function of the two Armenteros-Podolansky variables $p^{Arm}_{T}$ and $\alpha^{Arm}$~\cite{Armenteros-Podolanski} and after all track and topology selection criteria. The quantity $p^{Arm}_{T}$ is the transverse momentum of the decay particles with respect to the direction of motion of the $V^0$ candidate and $\alpha^{Arm} = (p^{+}_{L} - p^{−}_{L})/(p^{+}_{L} + p^{-}_{L})$, where $p^{+}_{L}$ and $p^{-}_{L}$ are the longitudinal momenta of the positively and negatively charged $V^0$ daughter particles, measured with respect to the $V^0$'s direction of motion. From the plots (see Fig.~\ref{fig:APplots}) one can see that contributions of $\Lambda$ and $\bar{\Lambda}$ hyperons are removed by the topological selection criteria.

\subsection{Raw \texorpdfstring{$K^0_S$}{Lg} yields}
\label{s:signal_extraction}
\begin{figure*}[h]
\centering
    \includegraphics[width=0.49\textwidth]{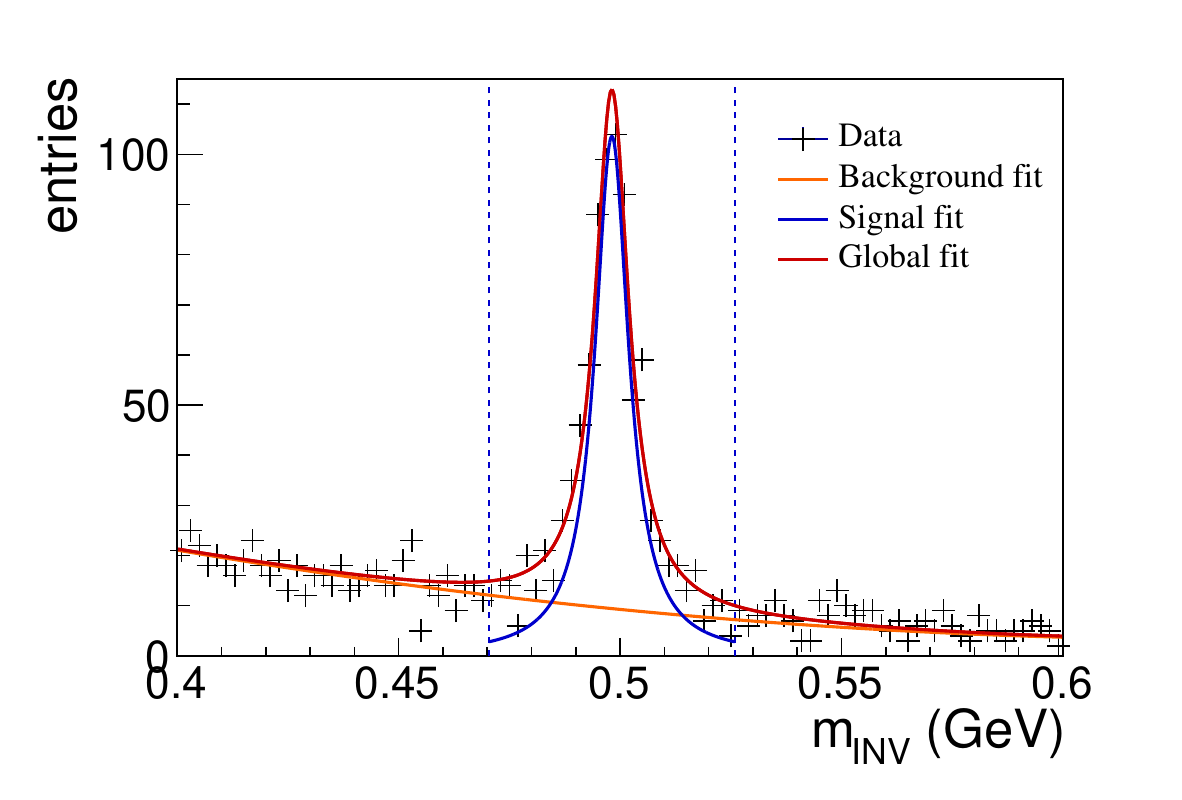}
    \includegraphics[width=0.49\textwidth]{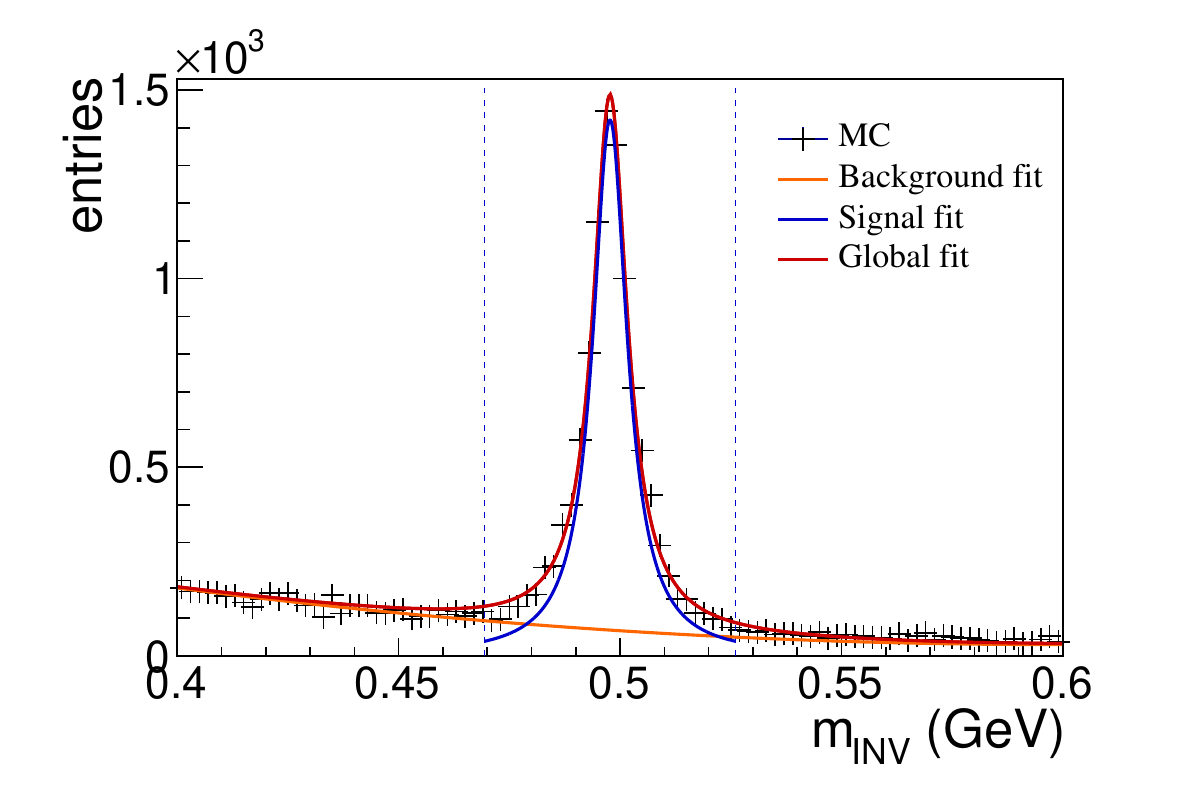} 
\caption[]{The invariant mass distribution of \Kshort candidates for experimental data (\textit{left}) and MC (\textit{right}) for the $p_{beam}=40$~\GeVc data set for $-0.25 \leq y < 0.25$ and $0.2 \leq p_{T} < 0.4$ after all selection criteria. The dashed-blue vertical lines indicate the regions where the \Kshort signal was integrated. The signal data points are black, the fitted background is orange, the fitted signal is blue, and the total fit results are red. Mass resolutions obtained from the fits are: $\sigma = (0.00925 \pm 0.00064)~\GeV$ for the experimental data and $\sigma = (0.00946 \pm 0.00017)~\GeV$ for the MC.
}
\label{fig:Fits}
\end{figure*}

The double differential uncorrected yields of \Kshort are determined by studying the invariant mass distributions of the accepted pion pairs in bins of rapidity and transverse momentum (examples are presented in Fig.~\ref{fig:Fits}). The \Kshort decays will appear as a peak over a smooth combinatorial background. The \Kshort yield was determined in each bin using a fit function that describes both the signal and the background. A Lorentzian function was used for the signal:

\begin{equation}
L(m) = A \frac{1}{\pi} \frac{\frac{1}{2}\Gamma}{(m-m_0)^2 + (\frac{1}{2}\Gamma)^2}~,
\end{equation}

where A is the normalization factor, $\Gamma$ is the full width at half maximum of the signal peak, and $m_0$ is the mass parameter. The background contribution is described by a polynomial function of $2^{nd}$ order. Figure~\ref{fig:Fits} shows examples of $\pi^{+} \pi^{-}$ invariant mass distributions obtained from the $p_{beam}=40$~\GeVc data set after all $V^0$ selection cuts for real data (\textit{left}) and for simulated events (\textit{right}). 

The procedure of fitting the histograms proceeds in three steps. In the first step, the background outside the signal peak ([0.475-0.525]~\GeV) is fitted with a polynomial of $2^{nd}$ order. This step is necessary to obtain starting values for the parameters of the background function. In the next step, a full invariant mass spectrum fit is performed with the sum of the Lorentzian and the background function. The initial parameter values for the background function are taken from the previous step, the mass parameter is fixed to the PDG value of $m_0 = 0.497614(24)$ \GeV~\cite{PDG}, and the width is allowed to vary between 0.005 and 0.03~\GeV. Finally, in the last step, all parameters are free, and the fitting region is [0.35-0.65]~\GeV.
The orange and blue curves in Fig.~\ref{fig:Fits} show the fitted polynomial background and the Lorentzian signal function. To minimize the sensitivity of the \Kshort yield to the integration window, the uncorrected number of \Kshort was calculated by subtracting bin-by-bin the fitted background (B) and summing the background-subtracted signal in the mass window $m_0 \pm 3\Gamma$ (dashed vertical lines), where $m_0$ is the fitted mass of the $K_S^0$.
Figure~\ref{fig:Fits} shows that the simulation reproduces the central value of the \Kshort mass distribution and its width agree with the data within uncertainties. The $\Gamma$ parameter fitted to the simulation was used to calculate the signal from the simulation. Thus, a possible bias due to differences between the data and the simulation is reduced; see Sec.~\ref{s:systematic_uncertainties}.

\begin{figure*}[h]
\centering
    \includegraphics[width=0.49\textwidth]{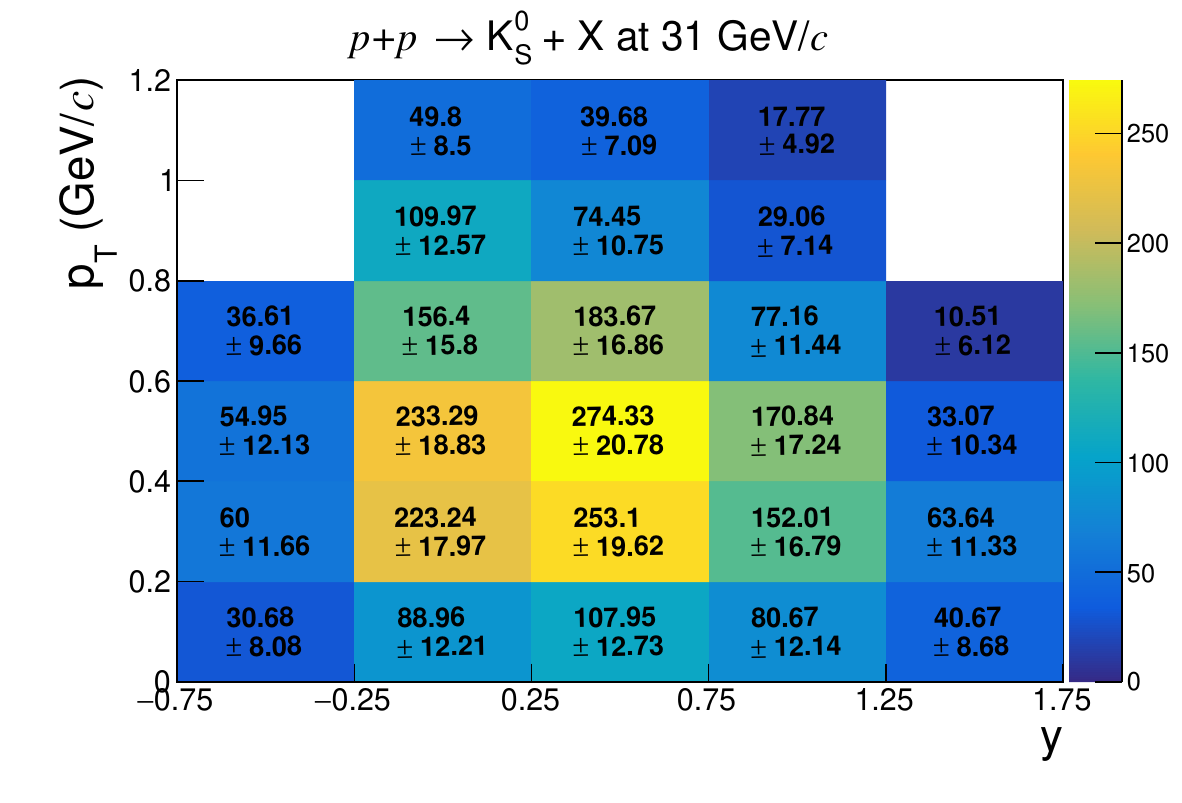}
    \includegraphics[width=0.49\textwidth]{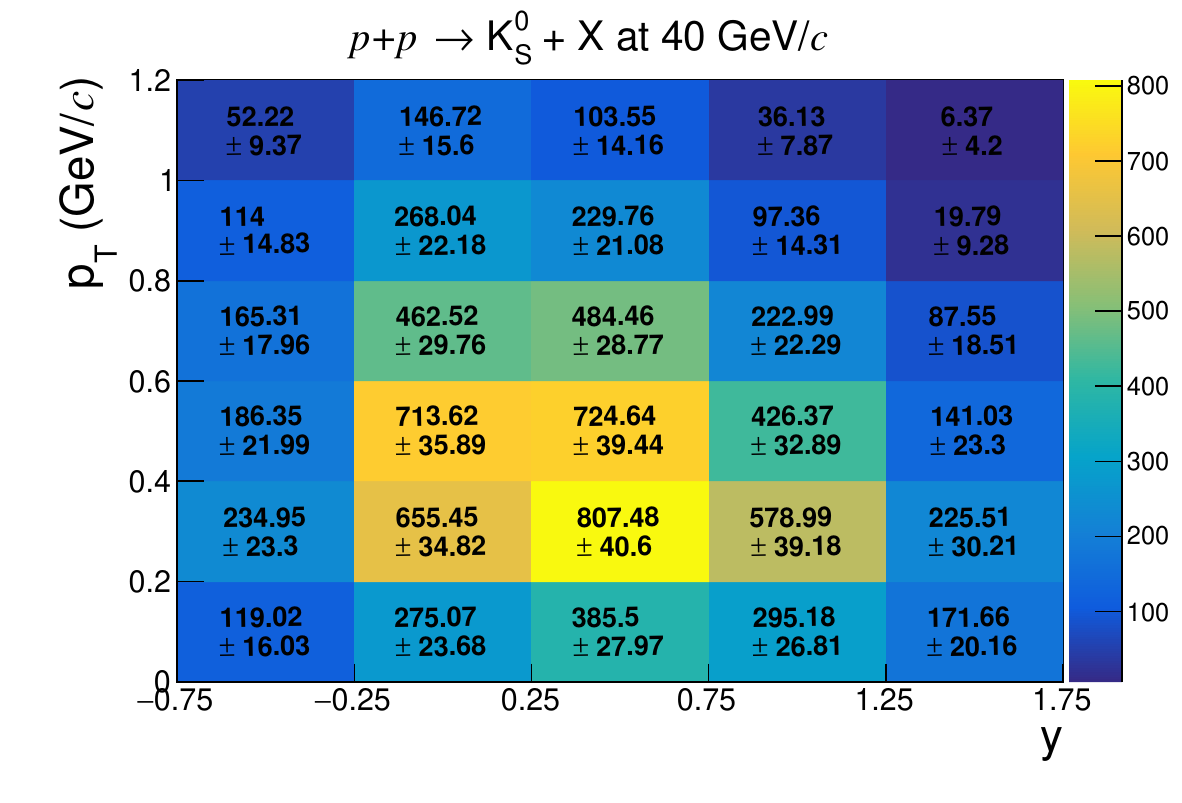} 
    \includegraphics[width=0.49\textwidth]{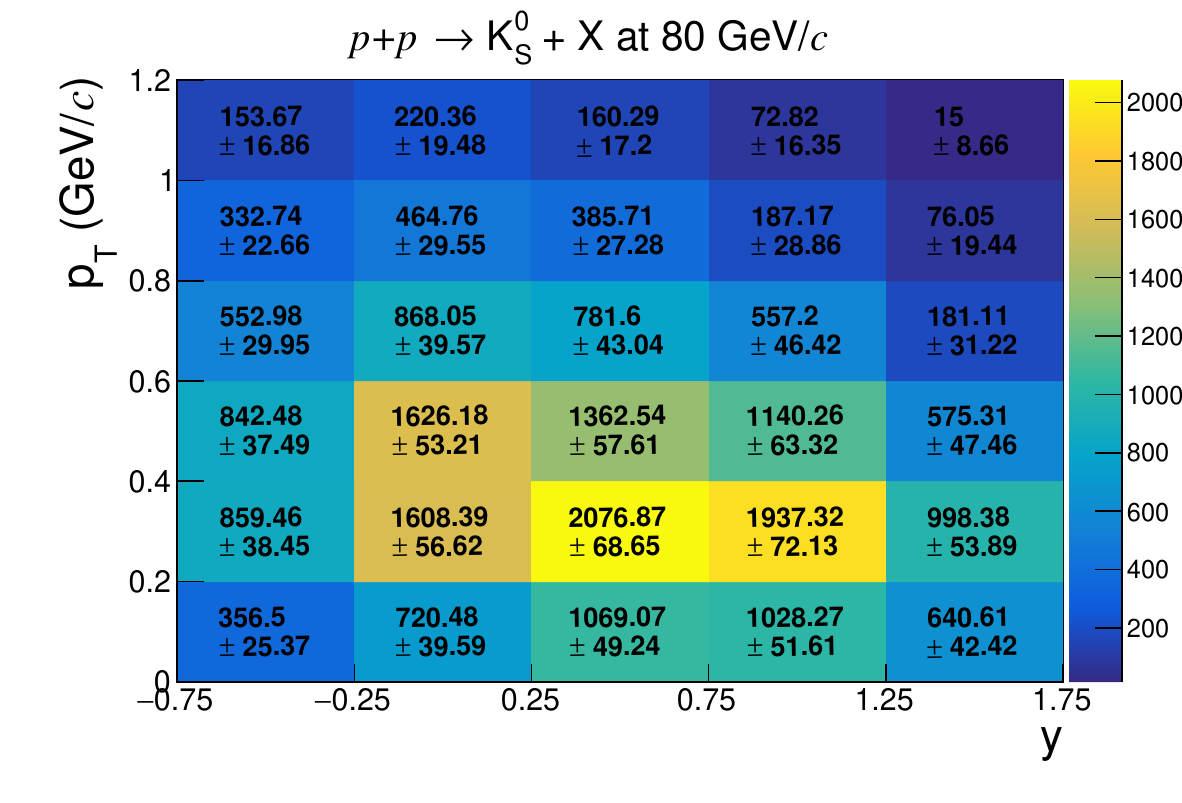}
    \caption[]{Uncorrected bin-by-bin multiplicities of \Kshort with their statistical uncertainties for $p_{beam}=31$~\GeVc (\textit{left}), $p_{beam}=40$~\GeVc (\textit{right}) and $p_{beam}=80$~\GeVc (\textit{bottom}).}
\label{fig:RawNo}
\end{figure*}

The uncorrected bin-by-bin \Kshort multiplicities and their statistical uncertainties are shown in Fig.~\ref{fig:RawNo}.

\subsection{Correction factors}
\label{sec:Correction_factors}
A correction for interactions of the incident protons with the target vessel is not needed, because the distributions of the primary vertex coordinates show no sign of such events after the event and track selection cuts.
A detailed Monte Carlo simulation was performed to compute the corrections for losses due to the trigger bias, geometrical acceptance, reconstruction efficiency, and the selection criteria applied in the analysis. The correction factors are based on $20 \times 10^6$ inelastic \textit{p+p} events at each beam momenta $p_{beam}=31, 40$ and $80$~\GeVc produced by the \EposLong event generator~\cite{Werner:2005jf, Pierog:2009zt}. Particles in the generated events were tracked through the \NASixtyOne apparatus using the \GeantThree package~\cite{GEANT}. The TPC response was simulated by dedicated software packages that account for known detector effects. The simulated events were reconstructed with the same software as the real events, and the same selection cuts were applied. However, \dedx identification was replaced by matching reconstructed tracks to simulated ones. The branching ratio of \Kshort decays is taken into account in the \GeantThree software package. For each $y$ and $p_T$ bin, the correction factor $c_{MC}(y,p_T)$ was calculated as:
	
\begin{equation}
    c_{MC} (y,p_T) =\left.\frac{n_{MC}^{gen}(y,p_T)}{N_{MC}^{gen}} \right/  \frac{n_{MC}^{acc}(y,p_T)}{N_{MC}^{acc}}~,
\label{eq:cmc}	
\end{equation}

where: 

\begin{itemize}
    \item [-] $n_{MC}^{gen}(y,p_T)$ is the number of \Kshort generated in a given ($y, p_T$) bin,
    \item [-] $n_{MC}^{acc}(y,p_T)$ is the number of  reconstructed \Kshort in a given ($y, p_T$) bin. 
    \item [-] $N_{MC}^{gen}$ is the number of generated inelastic \textit{p+p} interactions ($20 \times 10^6$),
    \item [-] $N_{MC}^{acc}$ is the number of accepted \textit{p+p} events (about $13.5 \times 10^6$ for all three beam momenta). 
\end{itemize}

The loss of the \Kshort mesons due to the \dedx cut is corrected with an additional factor:
\begin{equation}
    c_{dE/dx} = \frac{1}{\epsilon ^2} = 1.005~,
\end{equation}

where $\epsilon = 0.9973$ is the probability for the pions to be detected within $\pm 3\sigma$ around the nominal Bethe-Bloch value.	

The double-differential yield of \Kshort per inelastic event in bins of ($y, p_T$) is calculated as follows:

\begin{equation}
    \frac{d^2 n}{dy\, dp_T} (y, p_T) = \frac{c_{dE/dx} \cdot c_{MC}(y,p_T)}{\Delta y \, \Delta p_T} \cdot \frac{n_{K^{0}_{S}}(y,p_T)}{N_{events}}~,
\label{eq:dndydpt}
\end{equation}

where: 

\begin{itemize}
    \item [-] $ c_{dE/dx}$, $c_{MC}(y,p_T)$ are the correction factors described above,
    \item [-] $\Delta y$ and $\Delta p_T$ are the bin widths,
    \item [-] $n_{K^{0}_{S}}(y,p_T)$ is the uncorrected number of \Kshort, obtained by the signal extraction procedure described in Sec.~\ref{s:signal_extraction}. The corresponding values are presented in Fig.~\ref{fig:RawNo},
    \item [-] $N_{events}$ is the number of events left in the sample after selection criteria.
\end{itemize}

\subsection{Statistical uncertainties}
\label{s:statistical_uncertainties}
The statistical uncertainties of the corrected double-differential yields (see Eq.~\ref{eq:dndydpt}) receive contributions from the statistical uncertainty of the correction factor $c_{MC}(y,p_T)$ and the statistical uncertainty of the uncorrected number of \Kshort ($\Delta N_{K^{0}_{S}} (y, p_T)$). The statistical uncertainty of the former receives two contributions, the first, $\alpha$, caused by the loss of inelastic interactions due to the event selection and the second, $\beta$, connected with the loss of \Kshort candidates due to the $V^0$ selection:

\begin{equation}
    c_{MC} (y,p_T) =\left.\frac{n_{MC}^{gen}(y,p_T)}{N_{MC}^{gen}} \right/  \frac{n_{MC}^{acc}(y,p_T)}{N_{MC}^{acc}} = \left.\frac{N_{MC}^{acc}}{N_{MC}^{gen}} \right/  \frac{n_{MC}^{acc}(y,p_T)}{n_{MC}^{gen}(y,p_T)} = \frac{\alpha}{\beta(y,p_T)}~,
\end{equation}

The error of $\alpha$ is calculated assuming a binomial distribution:
\begin{equation}
    \Delta \alpha = \sqrt{\frac{\alpha(1-\alpha)}{N_{MC}^{gen}}}~,
\end{equation}

The error of $\beta$ is calculated according to the formula:
\begin{equation}
    \Delta \beta (y,p_T) = \sqrt{\left(\frac{\Delta n_{MC}^{acc}(y,p_T)}{n_{MC}^{gen}(y,p_T)} \right)^2+\left(\frac{n_{MC}^{acc}(y,p_T) \cdot \Delta n_{MC}^{gen}(y,p_T)}{(n_{MC}^{gen}(y,p_T))^2} \right)^2}~,
\end{equation}

where $\Delta n_{MC}^{acc}(y,p_T) = \sqrt{S+B}$ see Sec.~\ref{s:signal_extraction}, and $\Delta n_{MC}^{gen}(y,p_T)= \sqrt{n_{MC}^{gen}(y,p_T)}$. The equation for $\Delta c_{MC} (y,p_T)$ can be written as:

\begin{equation}
    \Delta c_{MC} (y,p_T) = \sqrt{\left(\frac{\Delta \alpha}{\beta}\right)^2+\left(-\frac{\alpha \cdot \Delta \beta}{\beta^2}\right)^2}~.
\label{eq:DMC}    
\end{equation}

\begin{figure*}[h]
\centering
    \includegraphics[width=0.49\textwidth]{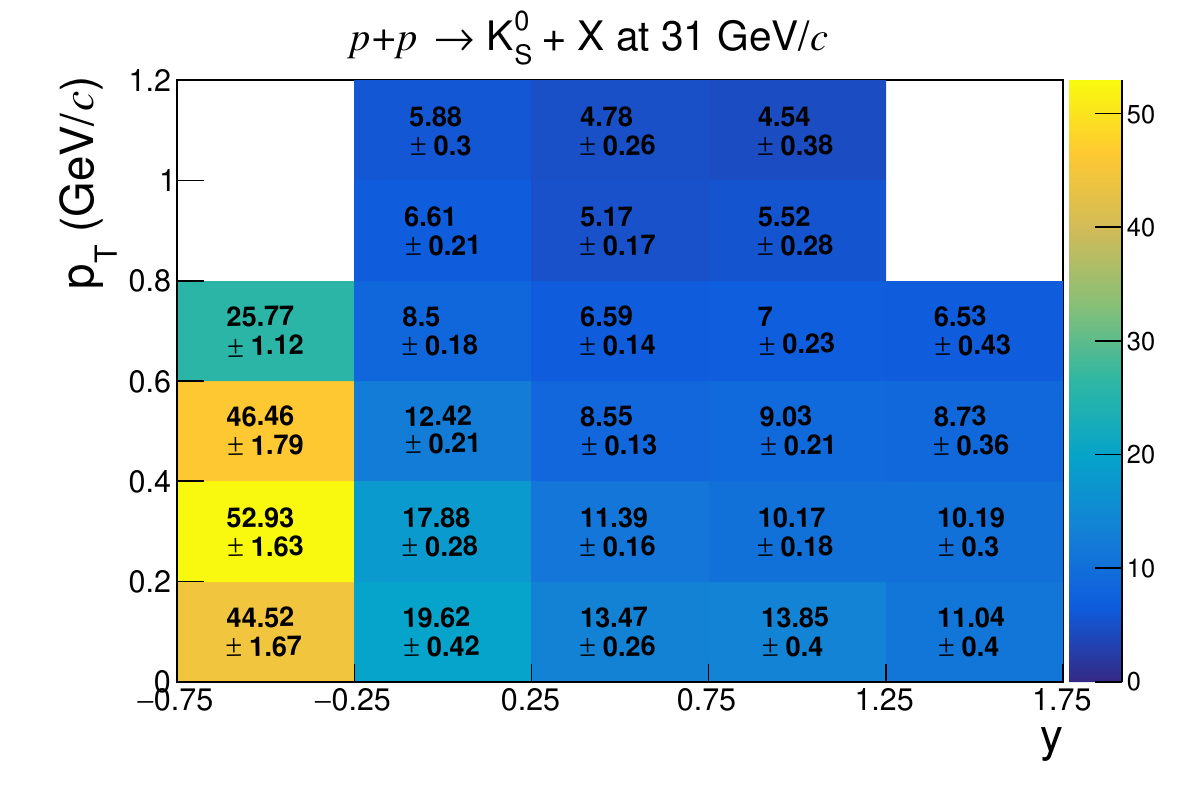}
    \includegraphics[width=0.49\textwidth]{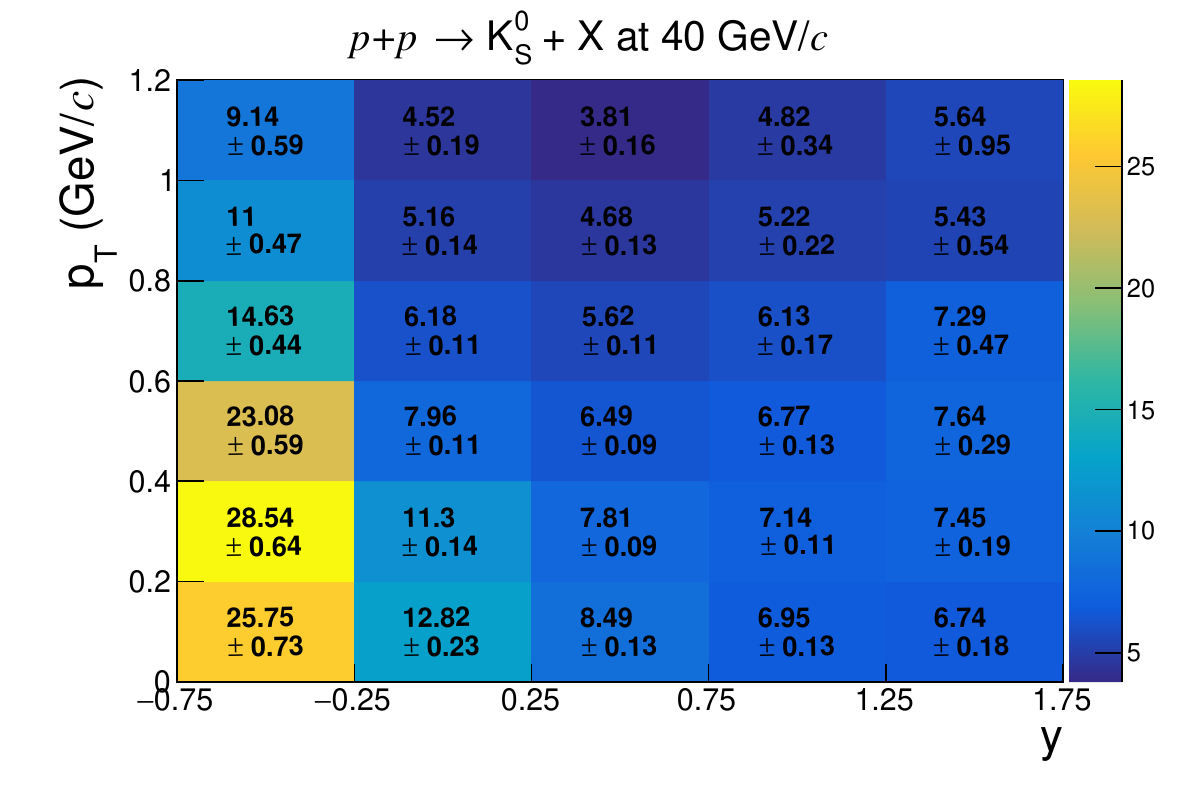} 
    \includegraphics[width=0.49\textwidth]{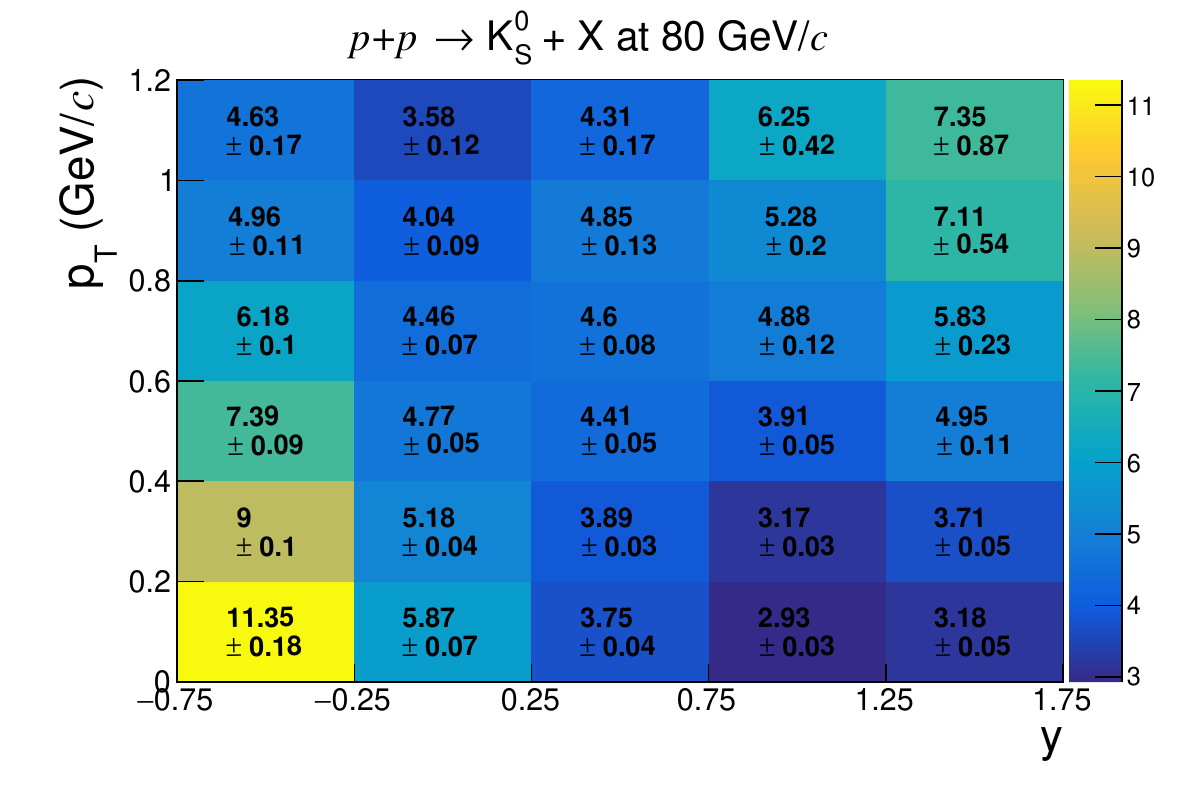}
\caption[]{Monte-Carlo correction factors (see Eq.~\ref{eq:cmc}) with their statistical uncertainties (see Eq.~\ref{eq:DMC}) in each ($y, p_T$) bin for $p_{beam}=31$~\GeVc (\textit{left}), $p_{beam}=40$~\GeVc (\textit{right}) and $p_{beam}=80$~\GeVc (\textit{bottom}).}
\label{fig:Cmc}
\end{figure*}

Finally, the statistical uncertainties $\Delta n_{K^{0}_{S}} (y, p_T)$ of the corrected number of \Kshort are:

\begin{equation}
    \Delta \frac{d^2 n}{dy dp_T} ( y, p_T ) =\sqrt{\left ( \frac{c_{dE/dx} \cdot c_{MC}(y,p_T)}{N_{events}\, \Delta y \, \Delta p_T} \right )^2 \Delta n_{K^{0}_{S}}^2 (y, p_T) + \left ( \frac{c_{dE/dx} \cdot n_{K^{0}_{S}}(y,p_T)}{N_{events}\, \Delta y \, \Delta p_T} \right)^2 \Delta c_{MC}^2(y, p_T)}~. 
\end{equation}

\subsection{Systematic uncertainties}
\label{s:systematic_uncertainties}
Three possible contributions to the systematic uncertainties related to the event selection criteria, the track and $V^0$ selection criteria and the signal extraction procedure were considered.

\begin{itemize}
    \item [(i)] The uncertainties related to the event selection criteria (see Sec.~\ref{s:event_selection}) were estimated by performing the analysis with the following changes:
	\begin{itemize}
        \item Simulations were done with and without the S4 trigger condition for all inelastic \textit{p+p} interactions. One-half of the difference between these two results was taken as the contribution to the systematic uncertainty, which amounts to  up to 3\%.
        \item The allowed range of the vertex \coordinate{z} position was changed from -590 < \coordinate{z}~(\cm) < -572 to -588 < \coordinate{z}~(\cm) < -574 and -592 < \coordinate{z}~(\cm) < -570. The uncertainty due to the variation of the selection window amounts to up to 4\%.
	\end{itemize}	
	
    \item [(ii)] The uncertainties related to the track and $V^0$ selection criteria were estimated by performing the analysis with the following changes compared to the original values (see Sec.~\ref{s:track_selection}):
	\begin{itemize}
        \item the minimum required number of clusters in both VTPCs for $V^0$ daughters was changed from 15 to 12 and 18, indicating a possible bias of up to 2\%,
        \item the standard \dedx cut used for identification of $V^0$ daughters was changed from $\pm 3\sigma$ to $\pm 2.5\sigma$ and $\pm 3.5\sigma$ from the nominal Bethe-Bloch value indicating a possible bias of up to 3\%,
        \item the $\Delta$\coordinate{z} cut was changed by varying the parameters $a$ and $b$ from 1.91 to 2.01 and 1.81 for parameter $a$ and from 0.99 to 0.98 and 1.00 for parameter $b$ for $p_{beam}=31$ \GeVc, from 1.71 to 1.91 and 1.51 for parameter $a$ and from 0.95 to 0.93 and 0.97 for parameter $b$ for $p_{beam}=40$ \GeVc and from 1.85 to 2.05 and 1.65 for parameter $a$ and from 0.90 to 0.88 and 0.92 for parameter $b$ for $p_{beam}=80$ \GeVc, indicating a possible bias of up to 2\%,
        \item the allowed distance of closest approach of the \Kshort trajectory to the primary vertex was varied from 0.25 to 0.20 and 0.30~cm, indicating a possible bias of up to 3\%, 
        \item the $cos\Theta^*$ range for accepted candidates was changed from $-0.97 < cos\Theta^* < 0.85$~to $-0.99 < cos\Theta^* < 0.87$~and $-0.95 < cos\Theta^* < 0.83$~indicating a possible bias of up to 3\%.
	\end{itemize}

    \item [(iii)] The uncertainty due to the signal extraction procedure (see Sec.~\ref{s:signal_extraction}) was estimated by:
	\begin{itemize}
        \item changing the background fit function from a $2^{nd}$ order to a $3^{rd}$ order polynomial indicating a possible bias of up to 4\%,
        \item changing the invariant mass range over which the uncorrected number of \Kshort was integrated from $m_0\pm 3\Gamma$ to $m_0\pm 2.5\Gamma$ and $m_0\pm 3.5\Gamma$ indicating a possible bias of up to 2\%, 
        \item calculating the uncorrected number of \Kshort as the sum of entries after background fit subtraction instead of the integral of the Lorentzian signal function indicating a possible bias of up to 2\%,
        \item changing the region of the fit from [0.35-0.65]~\GeVcc to [0.38-0.62]~\GeVcc~indicating a possible bias of up to 2\%.
    \end{itemize}

\end{itemize}

The maximum deviations are determined separately for each group of contributions to the systematic uncertainty. The systematic uncertainty was calculated as the square root of the sum of squares of the maximum deviations. This procedure was used to estimate systematic uncertainties of all final quantities presented in this paper: yields in ($y,p_T$) bins, inverse slope parameters of transverse momentum spectra, yields in rapidity bins, and mean multiplicities.

\subsection{Mean lifetime measurements}
\begin{figure*}[h]
\centering
    \includegraphics[width=0.49\textwidth]{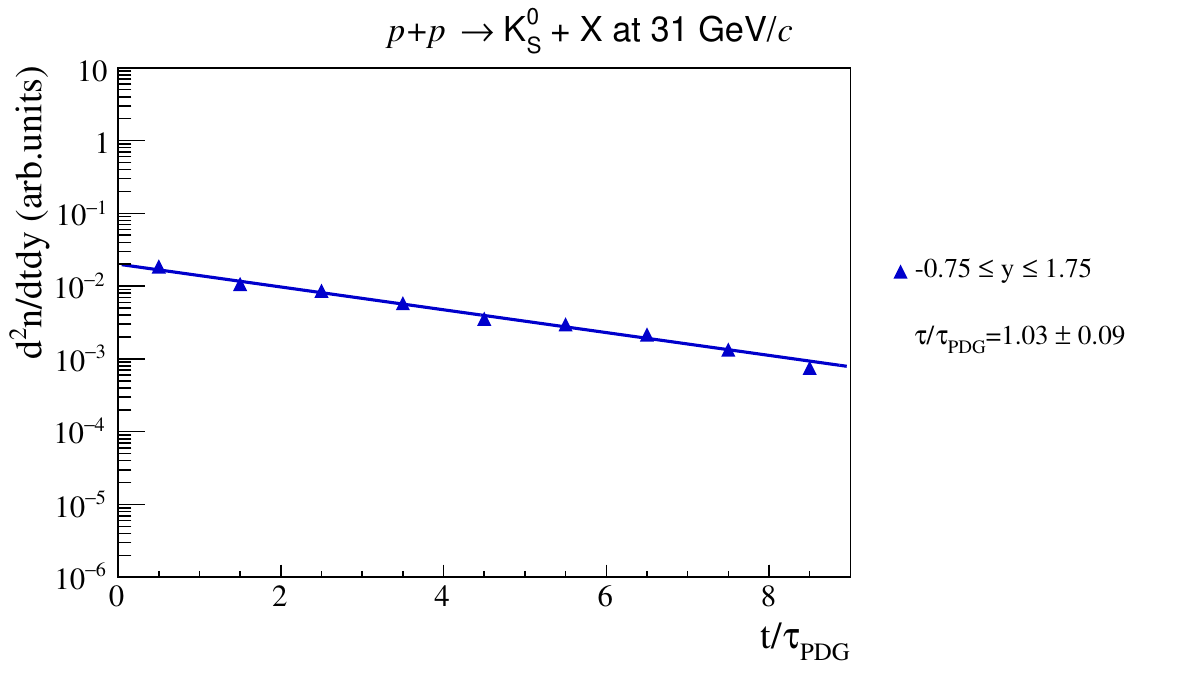} 
    \includegraphics[width=0.49\textwidth]{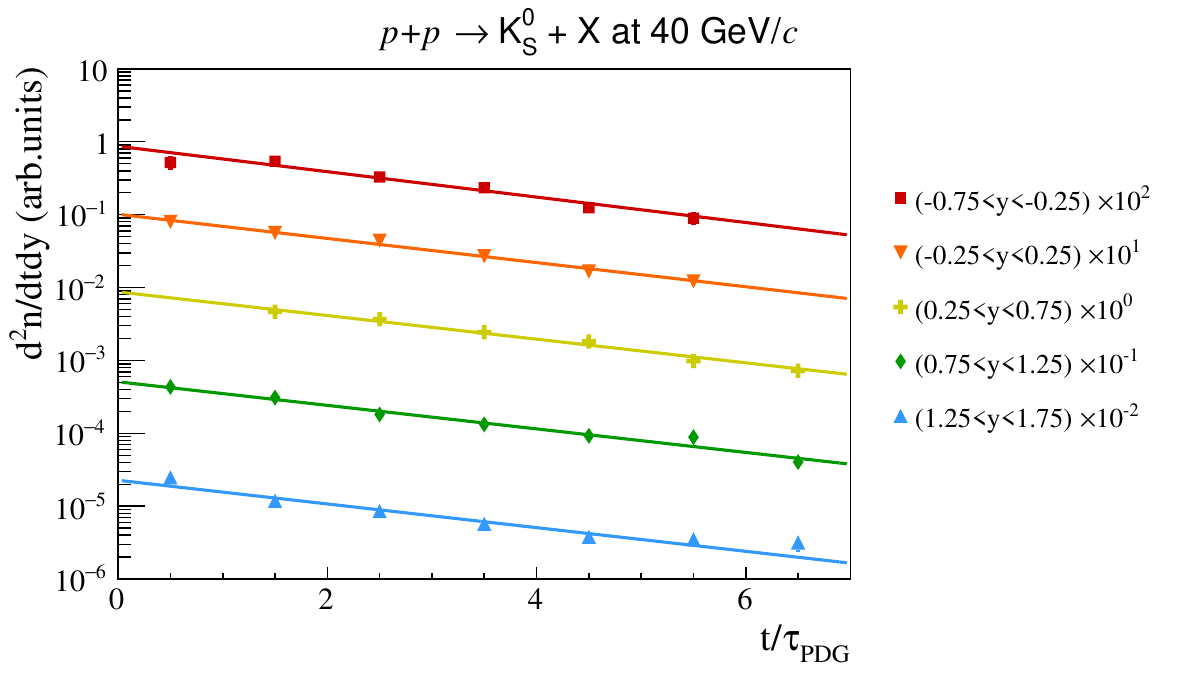}
    \includegraphics[width=0.49\textwidth]{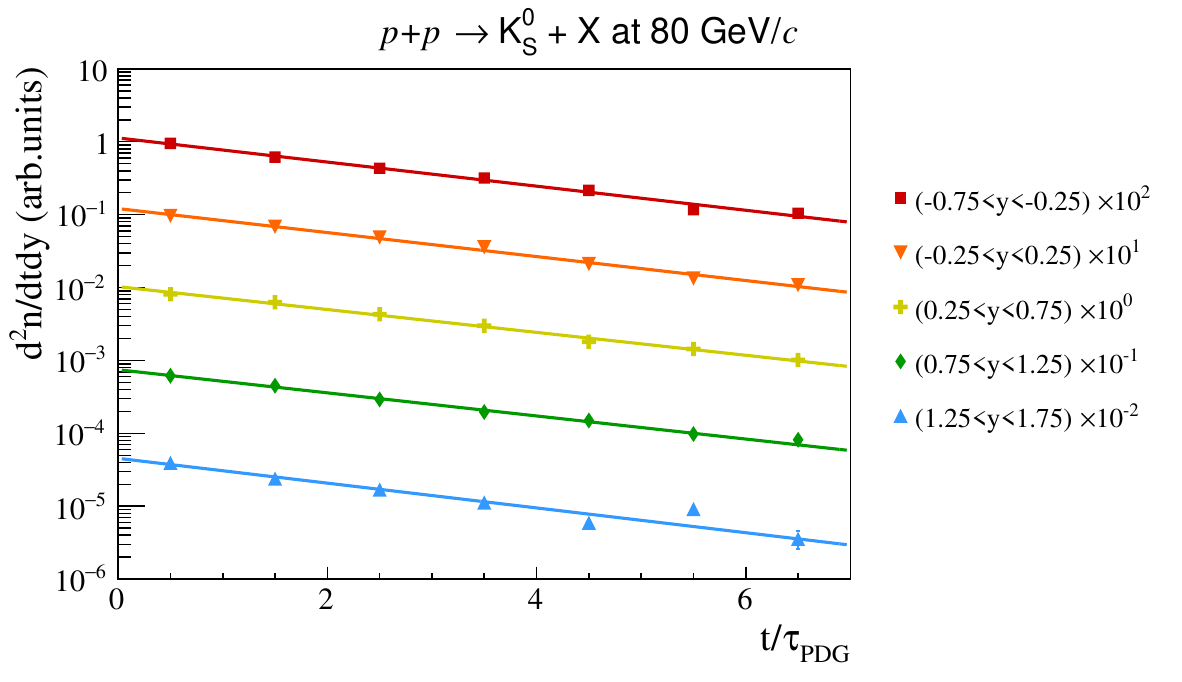}   
\caption[]{(Color online) Corrected lifetime distributions for \Kshort mesons produced in inelastic \textit{p+p} interactions at beam energies of $p_{beam}=31$~\GeVc (\textit{top left}), $p_{beam}=40$~\GeVc (\textit{top right}), and $p_{beam}=80$~\GeVc (\textit{bottom}). The straight lines show the results of exponential fits used to obtain the mean lifetimes (normalized to the PDG value) in rapidity bins. Statistical uncertainties are smaller than the marker size and are not visible on the plots.}
\label{fig:dndtau}
\end{figure*}

\begin{figure*}[h]
\centering
    \includegraphics[width=0.49\textwidth]{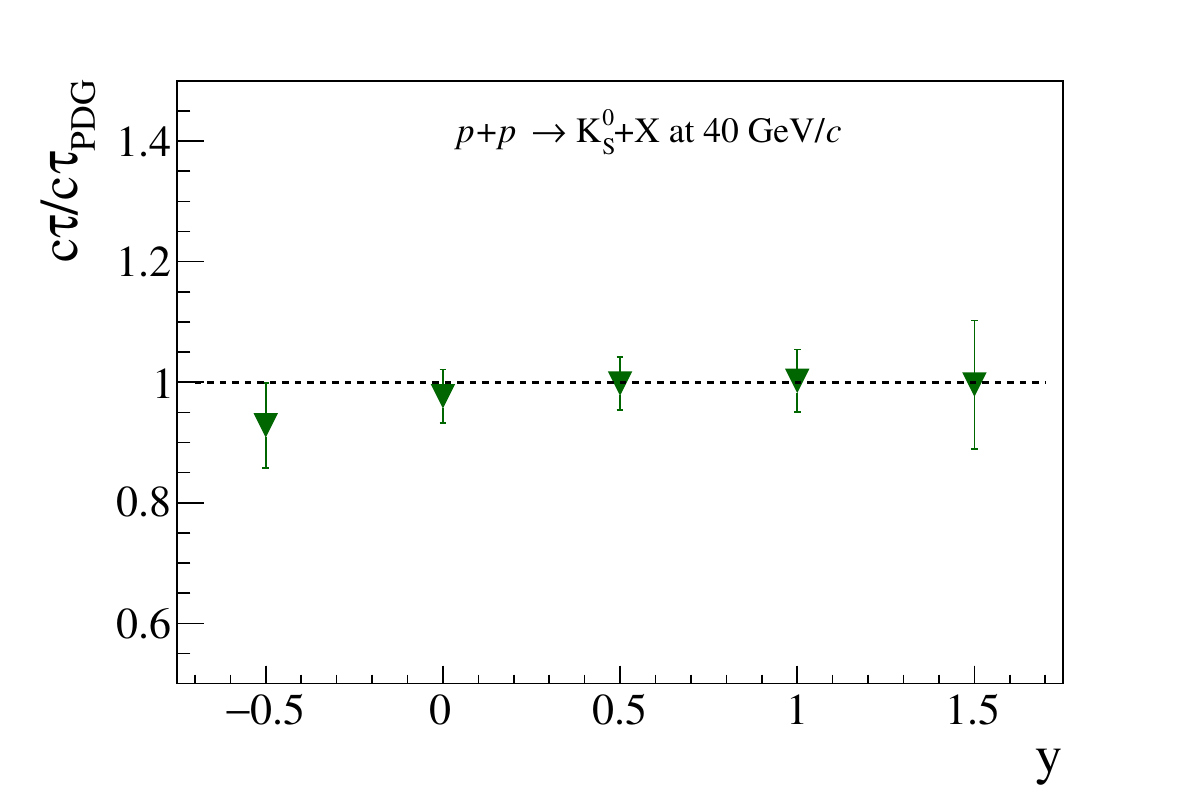}
    \includegraphics[width=0.49\textwidth]{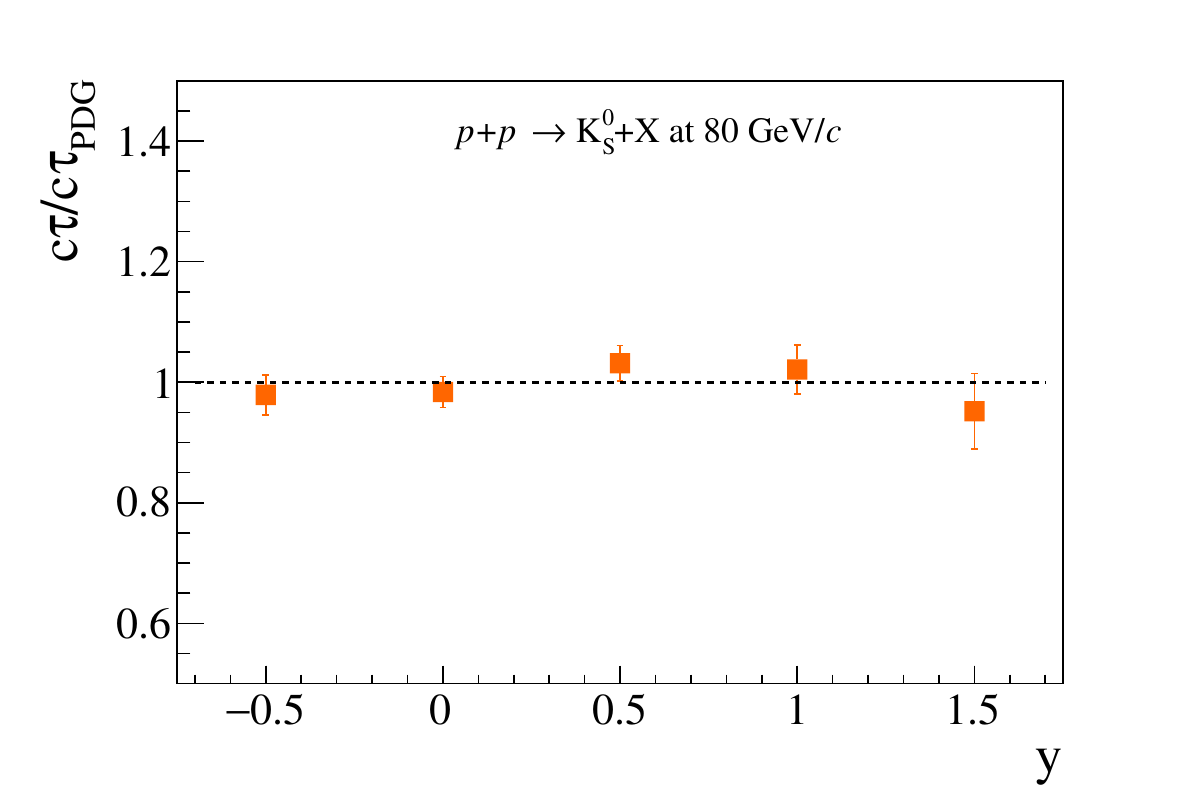}    
\caption[]{(Color online) Mean \Kshort lifetimes (normalized to the PDG value) obtained from fits to the lifetime distributions of Fig.~\ref{fig:dndtau} for the $p_{beam}=40$~\GeVc (\textit{left}) and $p_{beam}=80$~\GeVc (\textit{right}) data sets versus the rapidity $y$. The error bars indicate the statistical uncertainties.
}
\label{fig:lifetime}
\end{figure*}

The reliability of the \Kshort reconstruction and the correction procedure was validated by studying the lifetime distribution of the analyzed \Kshort. The lifetime ($c\tau$) of each identified \Kshort was calculated from the $V^0$ path length and its velocity. The corrected number of \Kshort was then determined in bins of $c\tau/c\tau_{PDG}$, and for the five rapidity bins of the $p_{beam}=40$ \GeVc and $80$ \GeVc data sets and in the whole rapidity range ($-0.75 < y < 1.75$) of the $p_{beam}=31$ \GeVc data set (see Fig.~\ref{fig:dndtau}). The straight lines in Fig.~\ref{fig:dndtau} represent the results of exponential fits, which provide mean lifetime values (normalized to the known PDG value~\cite{PDG}) as a function of rapidity. The thus determined mean lifetimes are shown in Fig.~\ref{fig:lifetime} as a function of rapidity. The measured mean \Kshort lifetimes agree within uncertainties with the PDG value and thus confirm the quality of the analysis. 

\FloatBarrier
\section{Results}
\label{sec:results}
This section presents new \NASixtyOne results on inclusive \Kshort meson production from inelastic \textit{p+p} interactions at beam momenta of 31, 40 and 80~\GeVc. Transverse momentum and rapidity spectra are obtained from the analysis of the weak decays of \Kshort mesons into two charged pions.

\subsection{Transverse momentum spectra}
\begin{figure*}
\centering
    \includegraphics[width=0.32\textwidth]{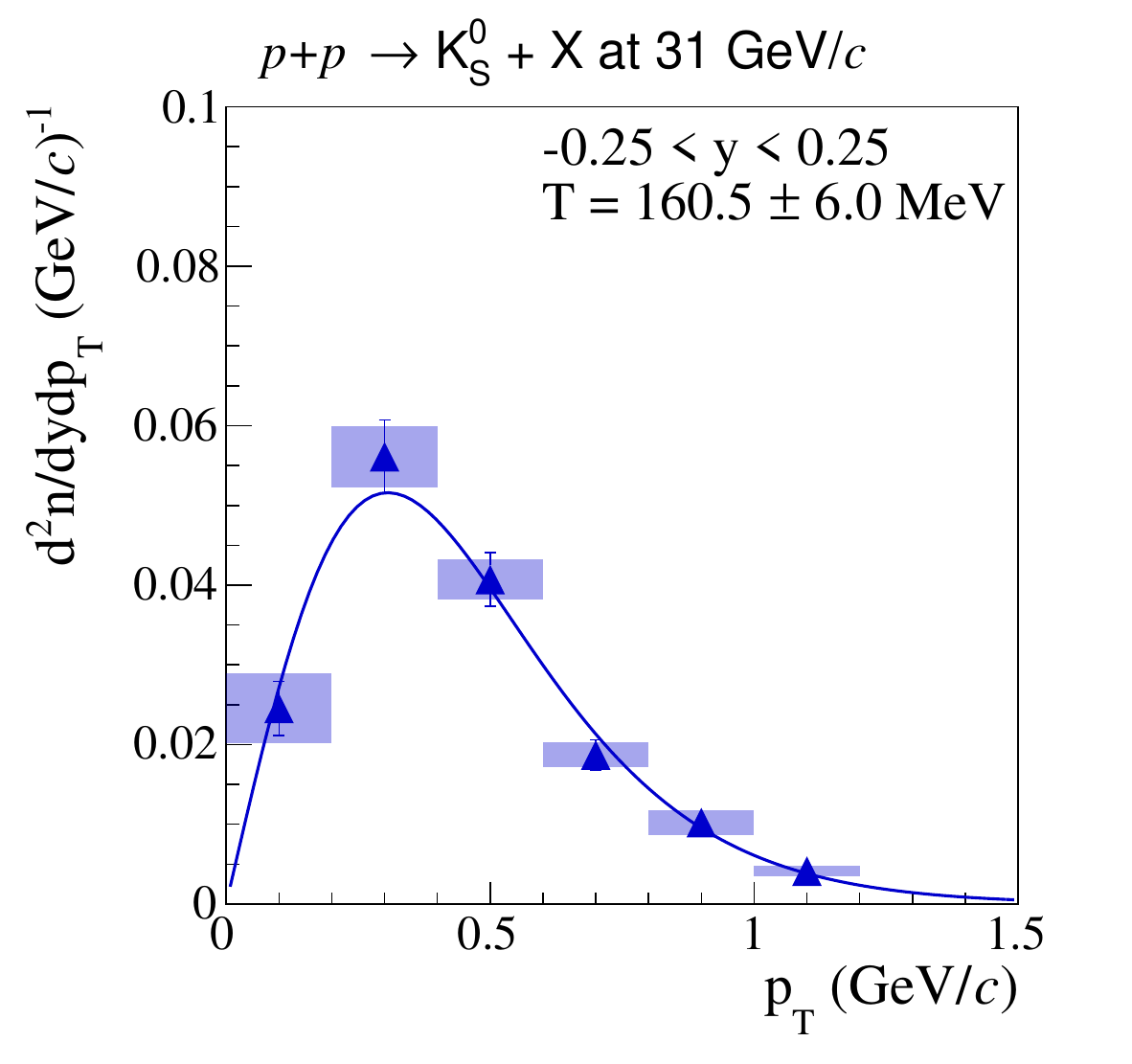}
    \includegraphics[width=0.32\textwidth]{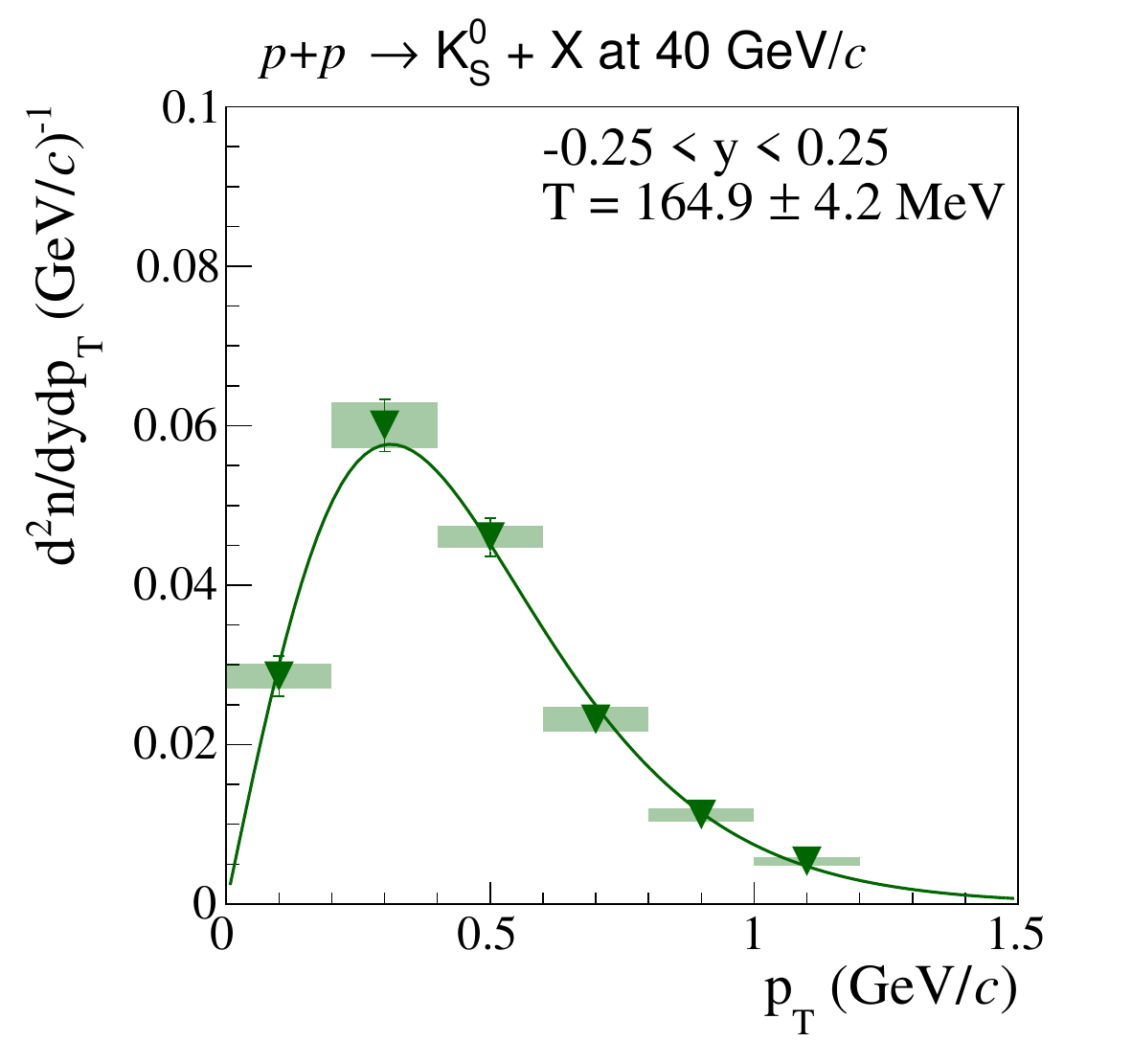}
    \includegraphics[width=0.32\textwidth]{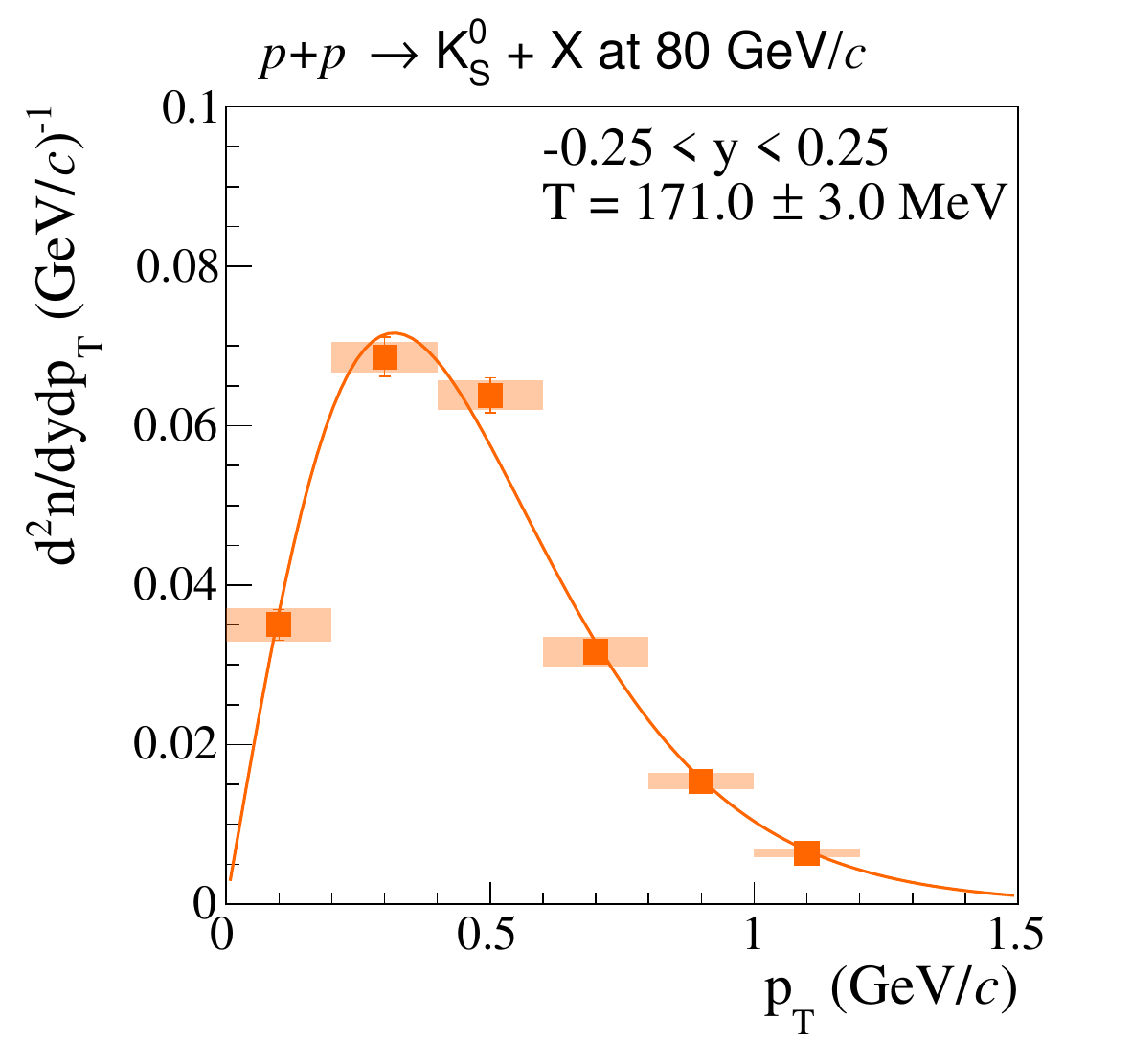}
\caption[]{(Color online) Double-differential \Kshort spectra in inelastic \textit{p+p} interaction at 31~\GeVc (\textit{left}), 40~\GeVc (\textit{middle}) and 80~\GeVc (\textit{right}) at mid-rapidity ($y \approx 0$) calculated according to Eq.~\ref{eq:dndydpt}. Measured points are shown as blue full triangles up (for $p_{beam}=31$~\GeVc), green full triangles down (for $p_{beam}=40$~\GeVc) and orange full squares (for $p_{beam}=80$~\GeVc). The solid curves are fitted to the data points using the exponential function (Eq.~\ref{eq:fit_to_dndydpt}). Vertical bars indicate statistical uncertainties (for some points smaller than the symbol size). Shaded boxes show systematic uncertainties. Only statistical uncertainties are taken into account in the fit, because the systematic uncertainties do not depend on \pt. The numerical values of the data points are listed in Table~\ref{tab:dndydpt}.}
\label{fig:sys_hpt}
\end{figure*}

Double differential \Kshort yields listed in Table~\ref{tab:dndydpt} represent the main result of this paper. Yields are determined in five consecutive rapidity bins in the interval $-0.75<y<1.75$ and six transverse momentum bins in the interval $0.0<p_T~(\GeVc)<1.2$. The transverse momentum distributions at mid-rapidity ($y \approx 0$) are shown in Fig.~\ref{fig:sys_hpt}.

\begin{table}[hbt!]
\centering
\small
\begin{tabular}{|c|c|c|c|c|}
    \hline
    Momentum &  & \multicolumn{3}{|c|}{Results $\frac{d^2n}{dydp_T} \times 10^{3}$} \\
    \hline
    \hline
    \cline{3-5}
    \multirow{12}{*}{$p_{beam}=31~\GeVc$} & \backslashbox{\hspace{1.1cm}$y$}{$p_T$~(\GeVc)} & (0.0;0.2) & (0.2;0.4) & (0.4;0.6) \\
    \cline{2-5}
    & (-0.75;-0.25) & 19.2 $\pm$ 5.1 $\pm$ 6.6 & 44.6 $\pm$ 8.8 $\pm$ 7.0 & 35.9 $\pm$ 8.0 $\pm$ 11.7 \\
    \cline{2-5}
    & (-0.25;0.25) & 24.5 $\pm$ 3.4 $\pm$ 4.4 & 56.1 $\pm$ 4.6 $\pm$ 3.9 & 40.7 $\pm$ 3.4 $\pm$ 2.6 \\
    \cline{2-5}
    & (0.25;0.75) & 20.4 $\pm$ 2.5 $\pm$ 1.7 & 40.5 $\pm$ 3.2 $\pm$ 2.0 & 33.0 $\pm$ 2.6 $\pm$ 2.4 \\
    \cline{2-5}
    & (0.75;1.25) & 15.7 $\pm$ 2.4 $\pm$ 4.6 & 21.7 $\pm$ 2.5 $\pm$ 1.9 & 21.7 $\pm$ 2.3 $\pm$ 2.1 \\
    \cline{2-5}
    & (1.25;1.75) & 6.3 $\pm$ 1.4 $\pm$ 2.5 & 9.1 $\pm$ 1.7 $\pm$ 1.2 & 4.1 $\pm$ 1.3 $\pm$ 1.0 \\
    \cline{2-5}
    \cline{3-5}
    &  \backslashbox{\hspace{1.1cm}$y$}{$p_T$~(\GeVc)} & (0.6;0.8) & (0.8;1.0) & (1.0;1.2) \\
    \cline{2-5}
    & (-0.75;-0.25) & 13.3 $\pm$ 3.6 $\pm$ 3.8 & - & - \\
    \cline{2-5}
    & (-0.25;0.25) & 18.7 $\pm$ 2.0 $\pm$ 1.6 & 10.2 $\pm$ 1.2 $\pm$ 1.6 & 4.2 $\pm$ 0.8 $\pm$ 0.7 \\
    \cline{2-5}
    & (0.25;0.75) & 17.0 $\pm$ 1.6 $\pm$ 1.3 & 5.4 $\pm$ 0.8 $\pm$ 0.6 & 2.7 $\pm$ 0.5 $\pm$ 0.4 \\
    \cline{2-5}
    & (0.75;1.25) & 7.6 $\pm$ 1.2 $\pm$ 0.5 & 2.3 $\pm$ 0.6 $\pm$ 0.7 & 1.2 $\pm$ 0.4 $\pm$ 0.2 \\
    \cline{2-5}
    & (1.25;1.75) & 1.0 $\pm$ 0.6 $\pm$ 0.3 & - & - \\
    \hline
    \hline
    \cline{3-5}
    \multirow{12}{*}{$p_{beam}=40~\GeVc$} & \backslashbox{\hspace{1.1cm}$y$}{$p_T$~(\GeVc)} & (0.0;0.2) & (0.2;0.4) & (0.4;0.6) \\
    \cline{2-5}
    & (-0.75;-0.25) & 24.8 $\pm$ 3.5 $\pm$ 7.3 & 54.3 $\pm$ 5.6 $\pm$ 7.6 & 34.9 $\pm$ 4.3 $\pm$ 4.3 \\
    \cline{2-5}
    & (-0.25;0.25) & 28.6 $\pm$ 2.6 $\pm$ 1.6 & 60.0 $\pm$ 3.3 $\pm$ 2.9 & 46.0 $\pm$ 2.4 $\pm$ 1.4 \\
    \cline{2-5}
    & (0.25;0.75) & 26.5 $\pm$ 2.0 $\pm$ 2.4 & 51.1 $\pm$ 2.7 $\pm$ 2.8 & 38.1 $\pm$ 2.2 $\pm$ 2.6 \\
    \cline{2-5}
    & (0.75;1.25) & 16.6 $\pm$ 1.6 $\pm$ 0.7 & 33.5 $\pm$ 2.4 $\pm$ 3.0 & 23.4 $\pm$ 1.9 $\pm$ 1.5 \\
    \cline{2-5}
    & (1.25;1.75) & 9.4 $\pm$ 1.2 $\pm$ 1.1 & 13.6 $\pm$ 1.9 $\pm$ 2.6 & 8.7 $\pm$ 1.5 $\pm$ 1.3 \\
    \cline{2-5}
    \cline{3-5}
    & \backslashbox{\hspace{1.1cm}$y$}{$p_T$~(\GeVc)} & (0.6;0.8) & (0.8;1.0) & (1.0;1.2) \\
    \cline{2-5}
    & (-0.75;-0.25) & 19.6 $\pm$ 2.3 $\pm$ 2.1 & 10.2 $\pm$ 1.4 $\pm$ 1.7 & 3.9 $\pm$ 0.8 $\pm$ 0.8 \\
    \cline{2-5}
    & (-0.25;0.25) & 23.2 $\pm$ 1.6 $\pm$ 1.6 & 11.2 $\pm$ 1.0 $\pm$ 0.9 & 5.4 $\pm$ 0.7 $\pm$ 0.6 \\
    \cline{2-5}
    & (0.25;0.75) & 22.1 $\pm$ 1.4 $\pm$ 1.8 & 8.7 $\pm$ 0.9 $\pm$ 1.2 & 3.2 $\pm$ 0.5 $\pm$ 0.4 \\
    \cline{2-5}
    & (0.75;1.25) & 11.1 $\pm$ 1.2 $\pm$ 1.0 & 4.1 $\pm$ 0.7 $\pm$ 0.5 & 1.4 $\pm$ 0.4 $\pm$ 0.3 \\
    \cline{2-5}
    & (1.25;1.75) & 5.2 $\pm$ 1.2 $\pm$ 2.3 & 0.9 $\pm$ 0.5 $\pm$ 0.2 & 0.3 $\pm$ 0.2 $\pm$ 0.1 \\
    \hline
    \hline
    \cline{3-5}
    \multirow{12}{*}{$p_{beam}=80~\GeVc$} & \backslashbox{\hspace{1.1cm}$y$}{$p_T$~(\GeVc)} & (0.0;0.2) & (0.2;0.4) & (0.4;0.6) \\
    \cline{2-5}
    & (-0.75;-0.25) & 35.2 $\pm$ 2.5 $\pm$ 3.5 & 64.2 $\pm$ 2.9 $\pm$ 2.6 & 51.8 $\pm$ 2.4 $\pm$ 2.9 \\
    \cline{2-5}
    & (-0.25;0.25) & 35.0 $\pm$ 2.0 $\pm$ 2.1 & 68.6 $\pm$ 2.5 $\pm$ 2.0 & 63.8 $\pm$ 2.2 $\pm$ 1.9 \\
    \cline{2-5}
    & (0.25;0.75) & 33.0 $\pm$ 1.6 $\pm$ 0.9 & 67.6 $\pm$ 2.3 $\pm$ 1.7 & 49.7 $\pm$ 2.2 $\pm$ 3.0 \\
    \cline{2-5}
    & (0.75;1.25) & 23.6 $\pm$ 1.3 $\pm$ 1.0 & 50.8 $\pm$ 2.0 $\pm$ 1.6 & 35.3 $\pm$ 2.1 $\pm$ 1.7 \\
    \cline{2-5}
    & (1.25;1.75) & 15.5 $\pm$ 1.2 $\pm$ 0.6 & 30.8 $\pm$ 1.7 $\pm$ 1.4 & 20.5 $\pm$ 1.9 $\pm$ 1.6 \\
    \cline{2-5}
    \cline{3-5}
    & \backslashbox{\hspace{1.1cm}$y$}{$p_T$~(\GeVc)} & (0.6;0.8) & (0.8;1.0) & (1.0;1.2) \\
    \cline{2-5}
    & (-0.75;-0.25) & 28.0 $\pm$ 1.6 $\pm$ 1.7 & 13.4 $\pm$ 1.0 $\pm$ 0.6 & 5.6 $\pm$ 0.7 $\pm$ 0.5 \\
    \cline{2-5}
    & (-0.25;0.25) & 31.6 $\pm$ 1.5 $\pm$ 1.9 & 15.4 $\pm$ 1.0 $\pm$ 1.1 & 6.4 $\pm$ 0.6 $\pm$ 0.5 \\
    \cline{2-5}
    & (0.25;0.75) & 29.3 $\pm$ 1.7 $\pm$ 1.6 & 14.7 $\pm$ 1.2 $\pm$ 1.0 & 5.4 $\pm$ 0.7 $\pm$ 0.4 \\
    \cline{2-5}
    & (0.75;1.25) & 22.0 $\pm$ 2.0 $\pm$ 1.3 & 7.4 $\pm$ 1.3 $\pm$ 0.9 & 3.4 $\pm$ 0.9 $\pm$ 0.3 \\
    \cline{2-5}
    & (1.25;1.75) & 8.5 $\pm$ 1.6 $\pm$ 1.5 & 4.6 $\pm$ 1.2 $\pm$ 1.3 & 0.8 $\pm$ 0.6 $\pm$ 0.2 \\
    \hline    
\end{tabular}

\caption{Double differential \Kshort yields in bins of ($y, p_T$). The first uncertainty is statistical, while the second one is systematic.
}
\label{tab:dndydpt}
\end{table}

An exponential function was fitted to the transverse momentum spectra. It reads:

\begin{equation}
f(p_T) = A \cdot p_T \cdot \exp \left(\frac{\sqrt{p_{T}^{2}+m_{0}^{2}}}{T}\right)~,
\label{eq:fit_to_dndydpt}
\end{equation}

where $m_{0}$ is the mass of the \Kshort and $T$ is the inverse slope parameter. The resulting values of $T$ in each rapidity bin are listed in Table~\ref{tab:dndy}.

\subsection{Rapidity distributions and mean multiplicities}
\label{sec:mean_multip}
\begin{figure*}[h]
\centering
    \includegraphics[width=0.7\textwidth]{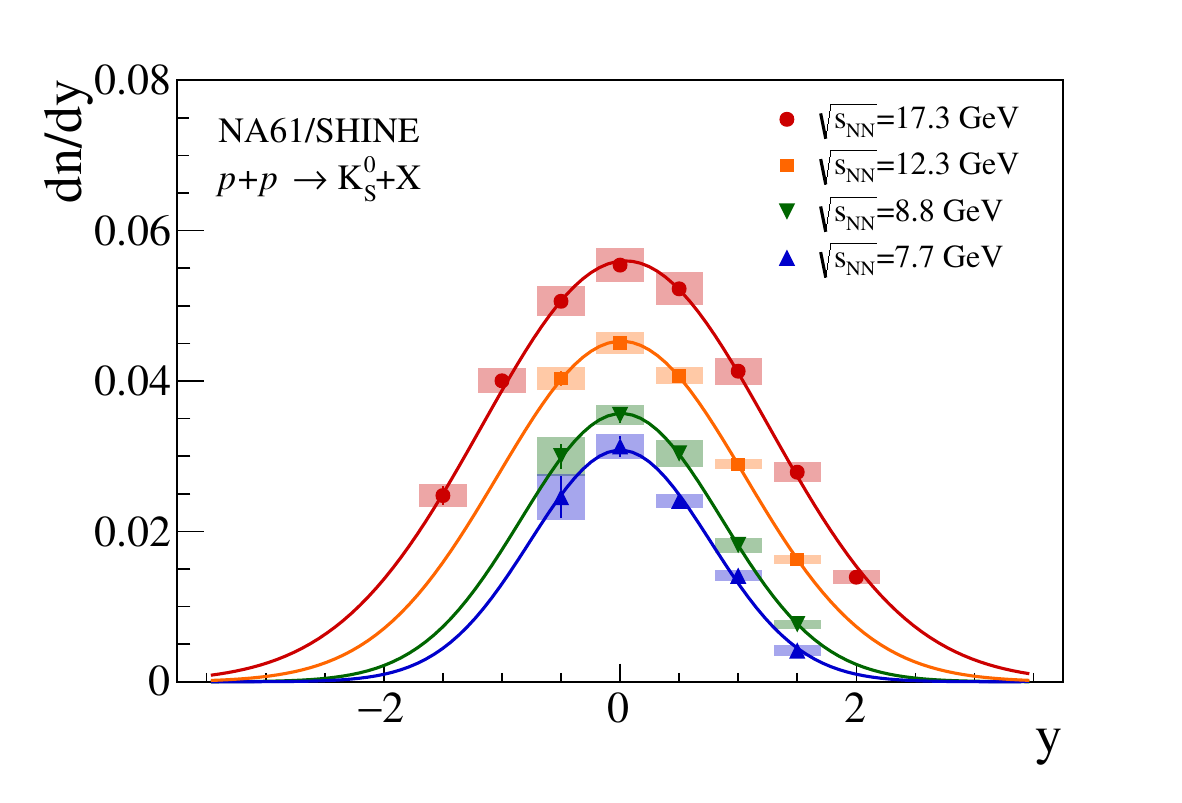}
\caption[]{(Color online) Rapidity distribution $dn/dy$ obtained by \pt-integration of data and extrapolation. Statistical uncertainties are shown by vertical bars (often smaller than the marker size), while shaded boxes indicate systematic uncertainties. The curves indicate the result of the Gaussian fit to the measured points. Points for \textit{p+p} at $\sqrt{s_{NN}}=17.3$ \GeV are taken from \cite{NA61SHINE:2021iay}.}
\label{fig:dndy}
\end{figure*}

Kaon yields in each rapidity bin were obtained from the measured transverse momentum distributions. The small fraction of \Kshort at high \pt outside of the acceptance was determined using Eq.~\ref{eq:fit_to_dndydpt}.
The resulting $\frac{dn}{dy}$ spectra of \Kshort mesons produced in inelastic \textit{p+p} interactions at 31, 40 and 80~\GeVc are presented in Fig.~\ref{fig:dndy} together with the previous \NASixtyOne results obtained for \textit{p+p} interactions at 158~\GeVc~\cite{NA61SHINE:2021iay}.

\begin{table}
\small
\centering
\begin{tabular}{|c|c|c|c|}
    \hline
    \multirow{6}{*}{$p_{beam}=31~\GeVc$} & $y$ & T (MeV) & $\frac{dn}{dy} \times 10^{3}$ \\
    \cline{2-4}
    & (-0.75;-0.25) & 149.4 $\pm$ 18.7 $\pm$ 21.8 & 24.6 $\pm$ 2.8 $\pm$ 3.1 \\
    \cline{2-4}
    & (-0.25;0.25) & 160.5 $\pm$ 5.9 $\pm$ 5.7 & 31.3 $\pm$ 1.5 $\pm$ 1.7 \\
    \cline{2-4}
    & (0.25;0.75) & 152.5 $\pm$ 4.9 $\pm$ 3.7 & 24.1 $\pm$ 1.1 $\pm$ 1.0 \\
    \cline{2-4}
    & (0.75;1.25) & 137.0 $\pm$ 6.2 $\pm$ 8.1 & 14.1 $\pm$ 0.9 $\pm$ 0.8 \\
    \cline{2-4}
    & (1.25;1.75) & 93.5 $\pm$ 11.7 $\pm$ 11.2 & 4.2 $\pm$ 0.6 $\pm$ 0.8 \\
    \hline
    \hline
    \multirow{6}{*}{$p_{beam}=40~\GeVc$} & $y$ & T (MeV) & $\frac{dn}{dy} \times 10^{3}$ \\
    \cline{2-4}
    & (-0.75;-0.25) & 162.6 $\pm$ 6.8 $\pm$ 9.7 & 30.0 $\pm$ 1.7 $\pm$ 2.6 \\
    \cline{2-4}
    & (-0.25;0.25) & 164.8 $\pm$ 4.2 $\pm$ 1.9 & 35.5 $\pm$ 1.0 $\pm$ 1.4 \\
    \cline{2-4}
    & (0.25;0.75) & 157.4 $\pm$ 3.7 $\pm$ 4.0 & 30.4 $\pm$ 0.9 $\pm$ 1.8 \\
    \cline{2-4}
    & (0.75;1.25) & 143.6 $\pm$ 4.5 $\pm$ 2.6 & 18.2 $\pm$ 0.8 $\pm$ 1.0 \\
    \cline{2-4}
    & (1.25;1.75) & 122.2 $\pm$ 7.6 $\pm$ 6.2 & 7.6 $\pm$ 0.6 $\pm$ 0.7 \\
    \hline
    \hline
    \multirow{6}{*}{$p_{beam}=80~\GeVc$} & $y$ & T (MeV) & $\frac{dn}{dy} \times 10^{3}$ \\
    \cline{2-4}
    & (-0.75;-0.25) & 165.8 $\pm$ 3.6 $\pm$ 2.3 & 40.3 $\pm$ 1.0 $\pm$ 1.6 \\
    \cline{2-4}
    & (-0.25;0.25) & 171.0 $\pm$ 3.0 $\pm$ 1.7 & 45.1 $\pm$ 0.9 $\pm$ 1.5 \\
    \cline{2-4}
    & (0.25;0.75) & 168.0 $\pm$ 3.4 $\pm$ 2.4 & 40.7 $\pm$ 0.8 $\pm$ 1.2 \\
    \cline{2-4}
    & (0.75;1.25) & 159.9 $\pm$ 4.7 $\pm$ 4.0 & 28.9 $\pm$ 0.8 $\pm$ 0.7 \\
    \cline{2-4}
    & (1.25;1.75) & 140.6 $\pm$ 6.2 $\pm$ 3.9 & 16.2 $\pm$ 0.7 $\pm$ 0.7 \\
    \hline    
    \end{tabular}
\vspace{0.5 cm}
\caption{Numerical values of $T$ and $dn/dy$ for \Kshort mesons produced in \textit{p+p} interactions at 31, 40 and 80~\GeVc. The first column indicates the data set. The second column shows the rapidity range. The values of the inverse slope parameter are listed in the third column, along with their statistical and systematic uncertainties. The last column shows the numerical values of the \pt-integrated yields presented in Fig.~\ref{fig:dndy} with statistical and systematic uncertainties. }

\label{tab:dndy}
\end{table}

The mean multiplicities of \Kshort mesons were calculated as the sum of the measured data points in Fig.~\ref{fig:dndy} scaled by the ratio between measured and unmeasured regions obtained from the Monte-Carlo simulation. The statistical uncertainties of $\langle K^{0}_{S} \rangle$ were calculated as the square root of the sum of the squares of the statistical uncertainties of the contributing bins. The systematic uncertainties were calculated as the square root of squares of systematic uncertainties described in Sec.~\ref{s:systematic_uncertainties}. To estimate the systematic uncertainties of the method used to determine the mean multiplicities of \Kshort, the rapidity distributions were also fitted using a single Gaussian or two Gaussians symmetrically displaced from mid-rapidity. The deviations of the results of these fits from $\langle K^{0}_{S} \rangle$ are included as an additional contribution to the final systematic uncertainty. The mean multiplicities of \Kshort mesons in inelastic \textit{p+p} collisions were found to be $0.0595 \pm 0.0019 (stat) \pm 0.0022 (sys)$ at 31~\GeVc, $0.0761 \pm 0.0013 (stat) \pm 0.0031 (sys)$ at 40~\GeVc and $0.1158 \pm 0.0012 (stat) \pm 0.0037 (sys)$ at 80~\GeVc.

\FloatBarrier
\section{Comparison with published world data and model calculations}
\label{sec:comparison}
\begin{figure*}[h]
\centering
    \includegraphics[width=0.70\textwidth]{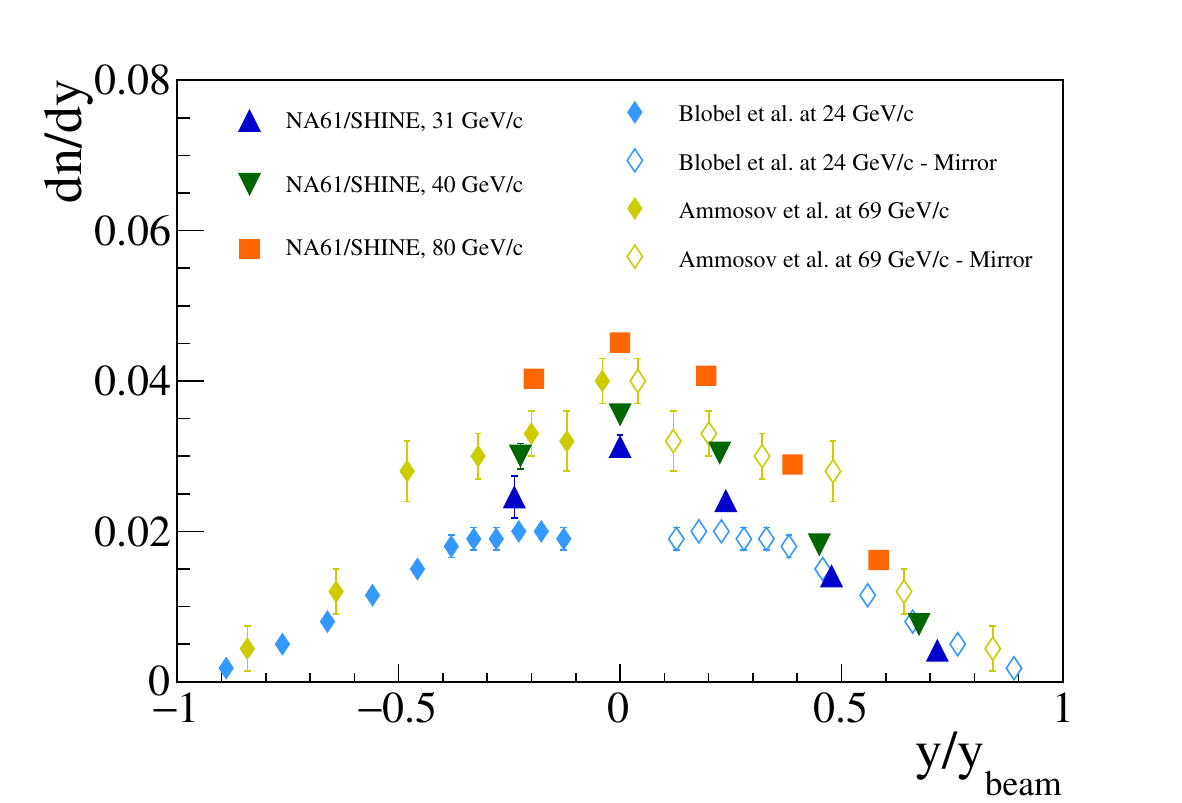}
\caption[]{(Color online) $dn/dy$ as a function of scaled rapidity $y/y_{beam}$ of \Kshort mesons in inelastic \textit{p+p} interactions at 31, 40 and 80~\GeVc. Measured points are shown as blue full triangles up (for $p_{beam}=31$~\GeVc), green full triangles down (for $p_{beam}=40$~\GeVc) and orange full squares (for $p_{beam}=80$~\GeVc). Results from other experiments are shown as azure-colored diamonds (for Blobel $et$ $al.$ at 24~\GeVc) and yellow-colored diamonds (for Ammosov $et$ $al.$ at 69~\GeVc). Vertical bars indicate statistical uncertainties (for some points smaller than the symbol size).}
\label{fig:dndy_comparison}
\end{figure*}

This section compares the new \NASixtyOne measurements of \Kshort production in inelastic \textit{p+p} interactions at 31, 40 and 80~\GeVc with world data as well as with microscopic model calculations (\EposLong~\cite{Werner:2005jf, Pierog:2009zt}, SMASH~2.0~\cite{Mohs:2019iee} and PHSD~\cite{Cassing:2008sv, Cassing:2009vt}).
The \Kshort rapidity spectra from \NASixtyOne are compared in Fig.~\ref{fig:dndy_comparison} to the results from Blobel $et$ $al.$~\cite{Blobel:1973jc} as well as with results from Ammosov $et$ $al.$~\cite{Ammosov:1975bt}. The results from Blobel $et$ $al.$ at 24 \GeVc~are significantly below the \NASixtyOne 31 \GeVc data in the central rapidity part. The results from Ammosov $et$ $al.$ at 69 \GeVc are located between the measured \NASixtyOne points of the 40 and 80~\GeVc data sets, as expected.

\begin{figure*}[h]
\centering
    \includegraphics[width=0.49\textwidth]{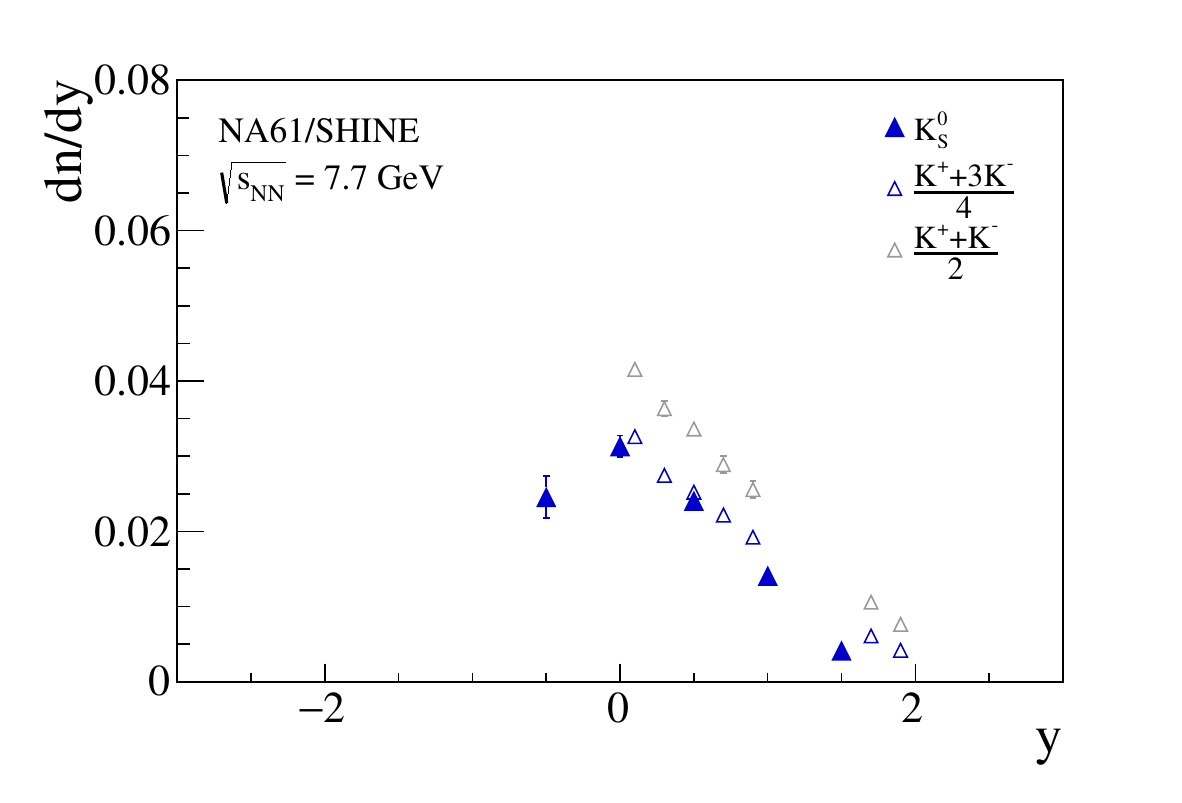}
    \includegraphics[width=0.49\textwidth]{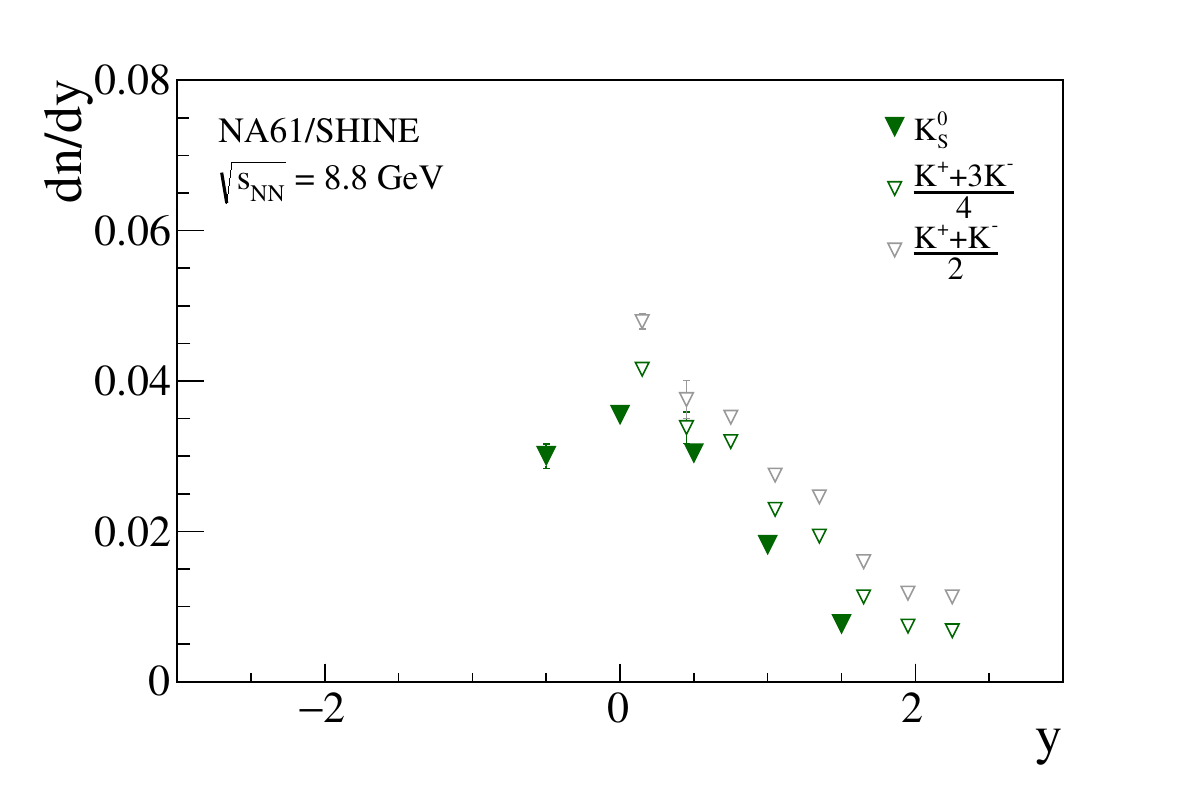}
    \includegraphics[width=0.49\textwidth]{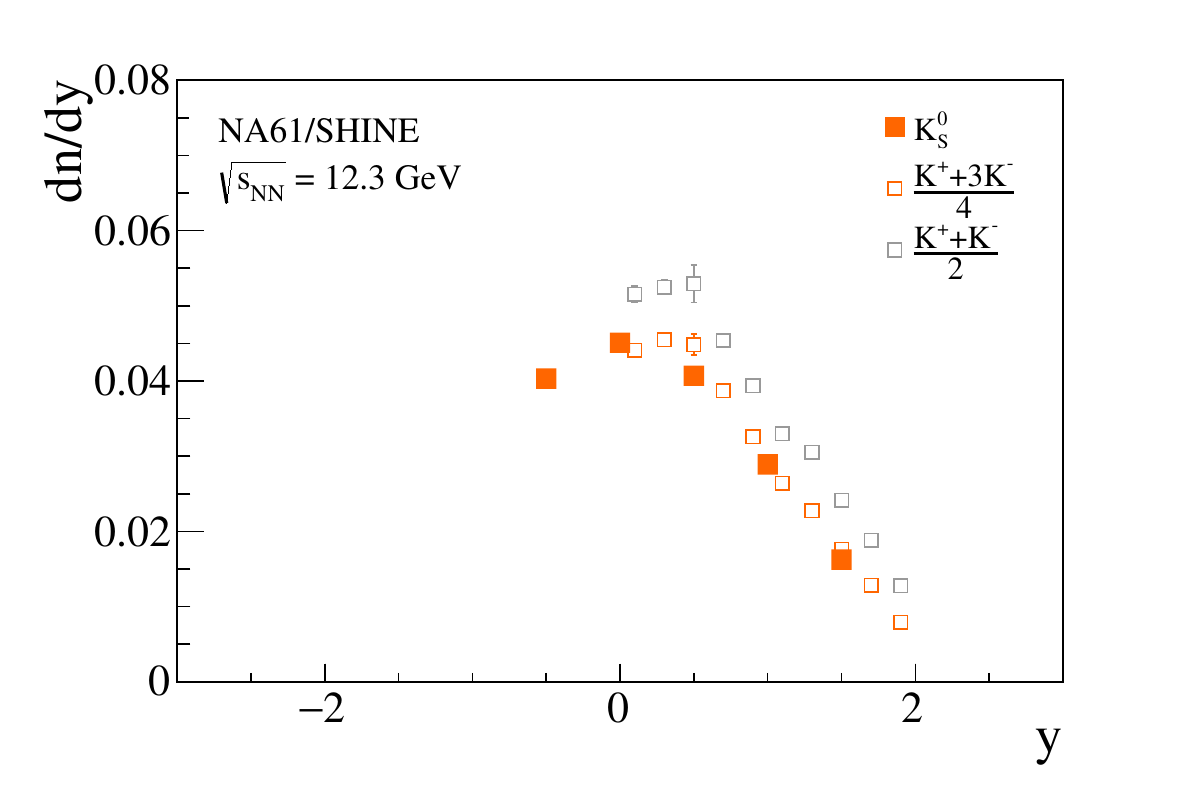} 
    \includegraphics[width=0.49\textwidth]{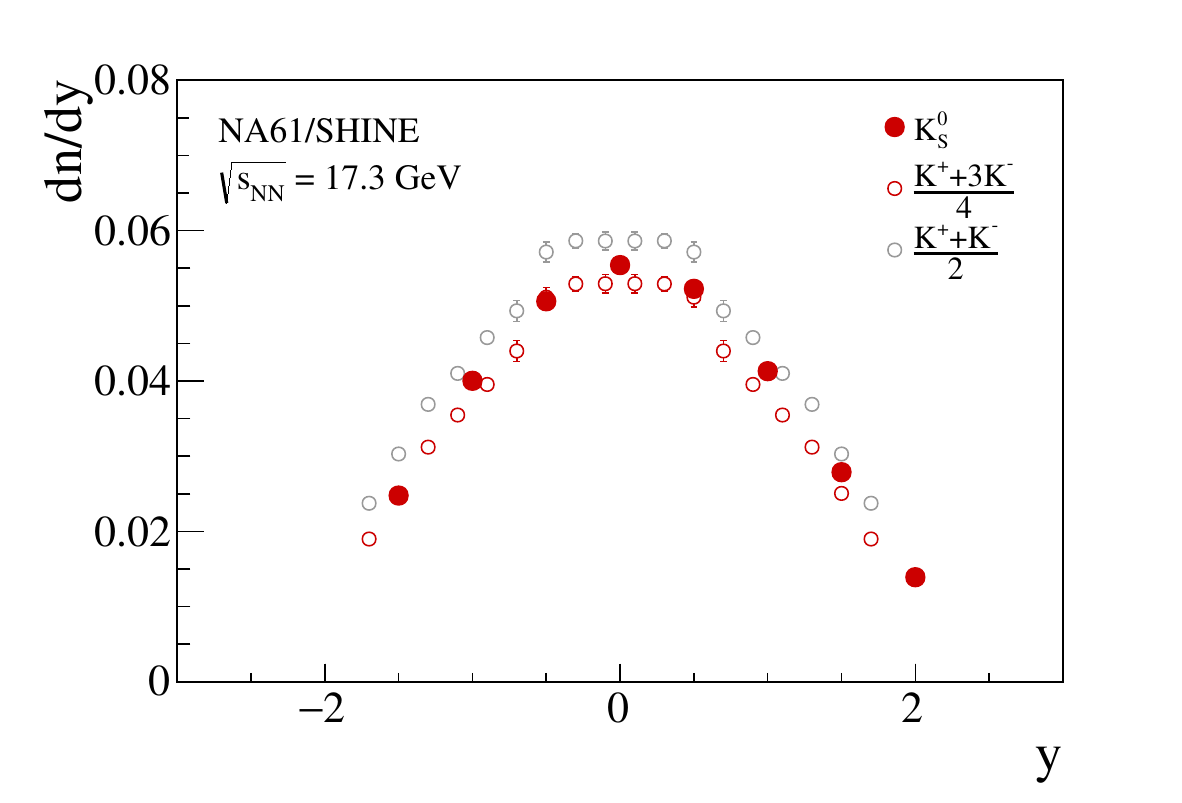}    
\caption[]{(Color online) Rapidity distribution $dn/dy$ of \Kshort mesons in inelastic \textit{p+p} interactions at 31, 40, 80 and 158~\GeVc. Measured points are shown as blue full triangles up for $p_{beam}=31$~\GeVc (\textit{top left}), green full triangles down for $p_{beam}=40$~\GeVc (\textit{top right}), orange full squares for $p_{beam}=80$~\GeVc (\textit{bottom left}) and red full circles for $p_{beam}=158$~\GeVc (\textit{bottom right}). Results for charged kaons obtained by formula $\frac{1}{4}(N_{K^+}+3 \cdot N_{K^-})$ are shown by open colored symbols for all data sets, while the results obtained by formula $\frac{1}{2}(N_{K^+} + N_{K^-})$ are shown by grey opened symbols. Vertical bars indicate statistical uncertainties (for some points smaller than the symbol size).
}
\label{fig:dndy_kaons}
\end{figure*}

Recently \NASixtyOne reported an excess of charged over neutral kaon production in Ar+Sc collisions at 75\textit{A} \GeVc~\cite{Adhikary:2883229}. The precise and detailed results on \Kshort production in \textit{p+p} interactions reported here, together with the corresponding results on charged kaons~\cite{NA61SHINE:2017fne}, may contribute to the understanding of this puzzle. To this end the rapidity distributions of \Kshort are compared with two predictions derived from $K^+$ and $K^-$ yields obtained from the same data sets~\cite{NA61SHINE:2017fne}. The first prediction is based on valence- and sea-quark counting arguments~\cite{Stepaniak:2023pvo} and leads to the equation $N_{K^0_S} = \frac{1}{4}(N_{K^+}+3 \cdot N_{K^-})$. This relation was used in the past to estimate the neutral kaon flux in the fragmentation region for $K^0$ beam studies~\cite{Gatignon:2730780}. The second prediction assumes isospin symmetry of the different charge states of the kaon: $N_{K^0_S} = \frac{1}{2}(N_{K^+}+N_{K^-})$. The \Kshort rapidity distributions are compared to these two predictions in Fig.~\ref{fig:dndy_kaons}. The prediction based on valence quark counting describes the \Kshort rapidity distributions significantly better than the one assuming isospin symmetry.

\begin{figure*}[h]
\centering
    \includegraphics[width=0.6\textwidth]{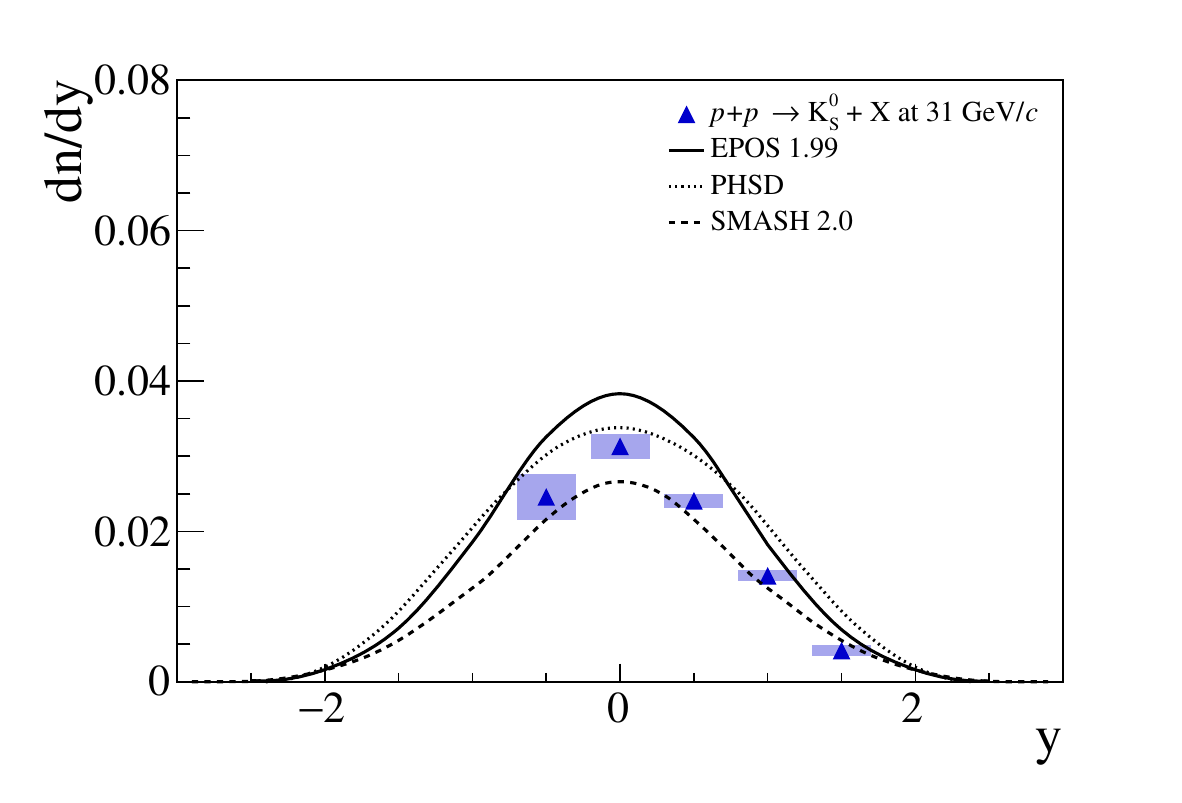}
    \includegraphics[width=0.6\textwidth]{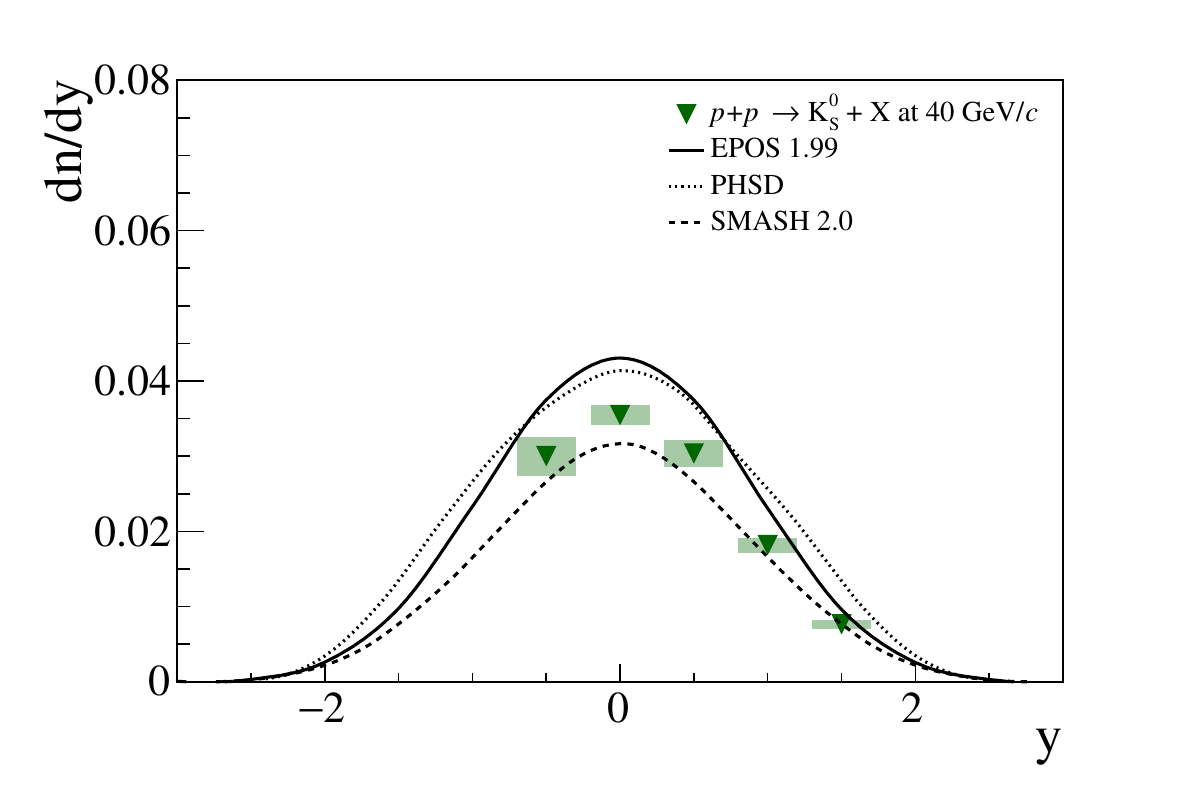}
    \includegraphics[width=0.6\textwidth]{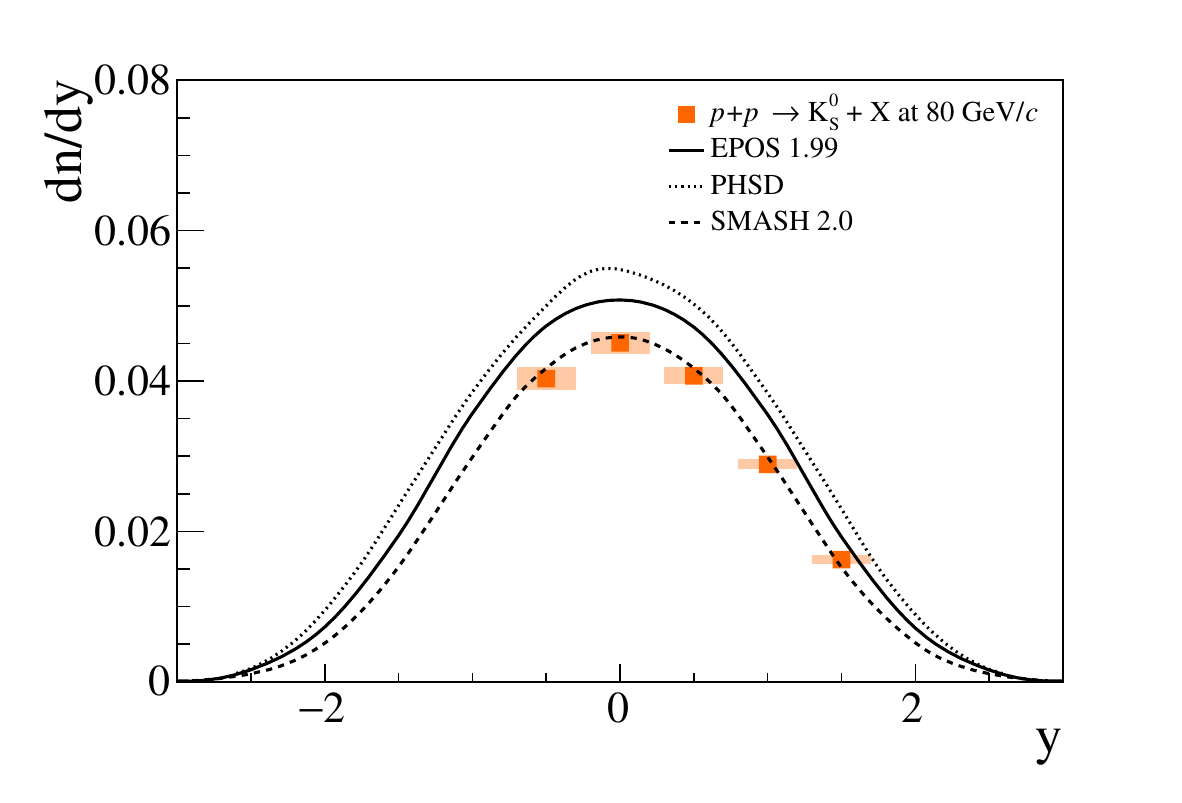}
\caption[]{(Color online) Comparison of the experimental \Kshort rapidity distributions with model calculations. Colored symbols show the new measurements of \NASixtyOne as follows: $p_{beam}=31$~\GeVc (\textit{top}), $p_{beam}=40$~\GeVc (\textit{middle}) and $p_{beam}=80$~\GeVc (\textit{bottom}). The black curves show the result of the model calculations: \EposLong (solid), PHSD (dotted) and SMASH~2.0 (dashed). }
\label{fig:models}
\end{figure*}

Figure~\ref{fig:models} compares the \NASixtyOne measurements with model calculations from \EposLong, PHSD and SMASH~2.0. \EposLong overpredicts the experimental data at all three data beam momenta. PHSD overpredicts the measured $p_{beam}=80$~\GeVc data, while for the remaining two data sets it shows fair agreement. SMASH~2.0 describes the experimental $p_{beam}=80$~\GeVc data very well but underpredicts the remaining two data sets. All models exhibit the same shape of the rapidity distribution as the experimental data. The energy dependence of \Kshort production seems to be well reproduced by \EposLong, whereas PHSD and SMASH~2.0 both exhibit a stronger rise than observed in the data.

\begin{figure*}[h]
\centering
    \includegraphics[width=0.7\textwidth]{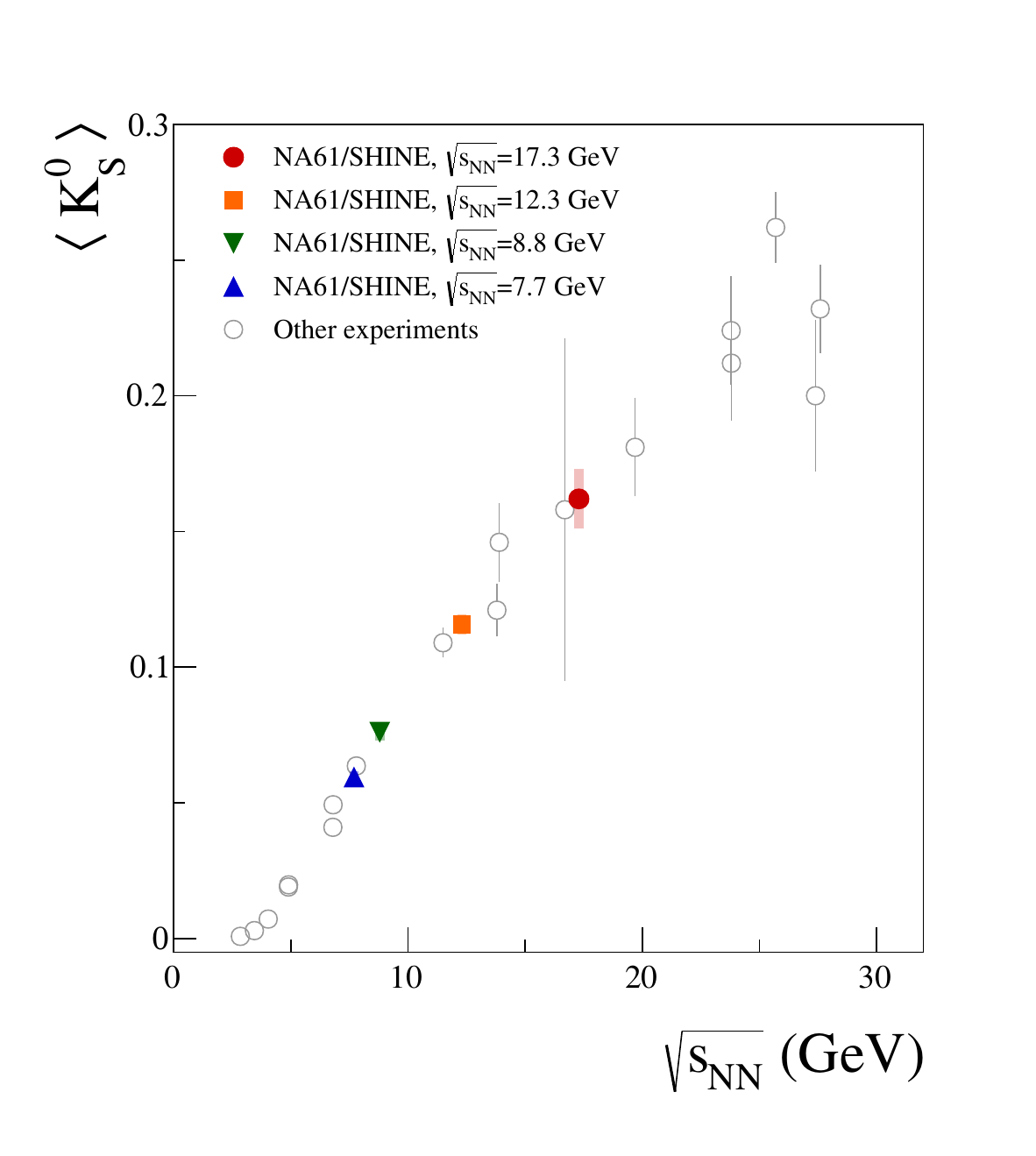}
\vspace{-0.5 cm}
\caption[]{(Color online) Collision energy dependence of mean multiplicity of \Kshort mesons produced in \textit{p+p} interactions. The measurements from \NASixtyOne are shown with colored symbols as follows: blue full triangle up for $p_{beam}=31$~\GeVc, green full triangle down for $p_{beam}=40$~\GeVc, orange full square for $p_{beam}=80$~\GeVc and red full circle for $p_{beam}=158$~\GeVc~\cite{NA61SHINE:2021iay}. The results published by other experiments are shown by the grey open circles~\cite{PhysRev.123.1465, Alexander:1967zz, Firebaugh:1968rq, Blobel:1973jc, Fesefeldt:1979, Bogolyubsky:1988ei, Ammosov:1975bt, AstonGarnjost:1975im, Chapman:1973fn, Brick:1980vj, Jaeger1974pk, Sheng1976, Lopinto:1980ct, EHS-RCBC:1984bxo, Kass:1979nf, Kichimi:1979te}. Statistical uncertainties are smaller than the marker size, while shaded boxes indicate systematic uncertainties.
}
\label{fig:multiplicity_comparison}
\end{figure*}

The mean multiplicity of \Kshort mesons in \textit{p+p} collisions at $\sqrt{s_{NN}}$ = 7.7, 8.8, 12.3~\GeV, reported here, and the published result at $\sqrt{s_{NN}}=17.3$~\GeV~\cite{NA61SHINE:2021iay} are compared in Fig.~\ref{fig:multiplicity_comparison} with the world data in the range from 3 - 32~\GeV. 
The measured values are seen to rise linearly with collision energy $\sqrt{s_{NN}}$.

\newpage
\FloatBarrier
\section{Summary}
\label{sec:summary}
This paper presents the new \NASixtyOne measurement of \Kshort mesons via their $\pi^{+} \pi^{-}$ decay mode in inelastic \textit{p+p} collisions at beam momenta of 31, 40 and 80~\GeVc ($\sqrt{s_{NN}}=7.7, 8.8$ and $17.3$~\GeV). Spectra of transverse momentum (up to 1.2 \GeVc), as well as a distributions of rapidity (from -0.75 to 1.75), are presented. The mean multiplicities, obtained from \pt-integrated spectra and extrapolated rapidity distributions, are $(5.95 \pm 0.19 \pm 0.22) \times 10^{-2}$ at 31~\GeVc, $(7.61 \pm 0.13 \pm 0.31) \times 10^{-2}$ at 40~\GeVc and $(11.58 \pm 0.12 \pm 0.37) \times 10^{-2}$ at 80~\GeVc, where the first uncertainty is statistical and the second systematic. 
The measured \Kshort lifetime agrees within uncertainties with the PDG value and thus confirms the quality of the analysis. 
The mean multiplicities from model calculations deviate by up to $10\%$ from the measurements. The SMASH~2.0 model provides the best results for $p_{beam}=31$ and $80$~\GeVc, while the PHSD model has the best agreement with measured data for $p_{beam}=40$~\GeVc. 
The results of \Kshort production in proton-proton interactions presented in this paper significantly improve, with their high statistical precision, the knowledge of strangeness production in elementary interactions and will serve as a reference for studies of strange hadron production in nucleus-nucleus collisions.


\FloatBarrier

\clearpage
\section*{Acknowledgments}
We would like to thank the CERN EP, BE, HSE and EN Departments for the
strong support of NA61/SHINE.

This work was supported by
the Hungarian Scientific Research Fund (grant NKFIH 138136\slash137812\slash138152 and TKP2021-NKTA-64),
the Polish Ministry of Science and Higher Education
(DIR\slash WK\slash\-2016\slash 2017\slash\-10-1, WUT ID-UB), the National Science Centre Poland (grants
2014\slash 14\slash E\slash ST2\slash 00018, 
2016\slash 21\slash D\slash ST2\slash 01983, 
2017\slash 25\slash N\slash ST2\slash 02575, 
2018\slash 29\slash N\slash ST2\slash 02595, 
2018\slash 30\slash A\slash ST2\slash 00226, 
2018\slash 31\slash G\slash ST2\slash 03910, 
2020\slash 39\slash O\slash ST2\slash 00277), 
the Norwegian Financial Mechanism 2014--2021 (grant 2019\slash 34\slash H\slash ST2\slash 00585),
the Polish Minister of Education and Science (contract No. 2021\slash WK\slash 10),
the European Union's Horizon 2020 research and innovation programme under grant agreement No. 871072,
the Ministry of Education, Culture, Sports,
Science and Tech\-no\-lo\-gy, Japan, Grant-in-Aid for Sci\-en\-ti\-fic
Research (grants 18071005, 19034011, 19740162, 20740160 and 20039012,22H04943),
the German Research Foundation DFG (grants GA\,1480\slash8-1 and project 426579465),
the Bulgarian Ministry of Education and Science within the National
Roadmap for Research Infrastructures 2020--2027, contract No. D01-374/18.12.2020,
Serbian Ministry of Science, Technological Development and Innovation (grant
OI171002), Swiss Nationalfonds Foundation (grant 200020\-117913/1),
ETH Research Grant TH-01\,07-3, National Science Foundation grant
PHY-2013228 and the Fermi National Accelerator Laboratory (Fermilab),
a U.S. Department of Energy, Office of Science, HEP User Facility
managed by Fermi Research Alliance, LLC (FRA), acting under Contract
No. DE-AC02-07CH11359 and the IN2P3-CNRS (France).\\

The data used in this paper were collected before February 2022.

\clearpage

\bibliographystyle{include/na61Utphys}
{\footnotesize\raggedright
\bibliography{include/na61References}
}

\end{document}